\documentclass{article}
\usepackage[english,frenchb]{babel}
\usepackage[pdftex]{graphicx}
\usepackage{ifthen}
\title{A  rational parametrization of Bezier like curves}
\author{
Mohamed ALLAOUI\footnote{Unit\'e de 
Math\'ematiques IMSP/UP} and 
Aur\'elien GOUDJO\footnote{D\'epartement de 
Math\'ematiques FAST/UAC, Email : aurelien.goudjo@uac.bj}\\
}
\date{}
\newcommand{\RR}{{\bf I\!\!R}}
\newcommand{\NN}{{\bf I\!\!\!N}}

\newcommand{\norme}[1]{\left\Vert {#1} \right\Vert}
\newcommand{\suite}[1]{{\left( {#1}  \right)}}
\newcommand{\couple}[2]{{\left( {#1} \, , \, {#2} \right)}}

\newcommand{\implique}{ \, \Rightarrow \,}
\newcommand{\equivaut}{ \, \Leftrightarrow \,}

\newcommand{\module}[1]{\left\vert {#1} \right\vert}
\newcommand{\Preuve}{\noindent{\bf \underline{Proof} } \\}

\newcommand{\Bezier}[3]{{{}^{#3}}{\bf B}_{#1}^{#2}}

\newcommand{\osegment}[2]{{\left({#1}\, , \, {#2}\right)}}

\newtheorem{definition}{\bf Definition}[section]

\newtheorem{lemme}{\bf Lemma}[section]

\newtheorem{proposition}{\bf Proposition}[section]
\newtheorem{corollaire}{\bf Corollary}[section]
\newtheorem{remarque}{\bf Remark}[section]

\newtheorem{castest}{\bf Illustration}[section]

\newboolean{isthese}
\setboolean{isthese}{false}
\begin{document}
\maketitle

\begin{abstract}
Une classe de param\'etrisation rationnelle a \'et\'e d\'evelopp\'ee puis utilis\'ee 
pour g\'en\'erer des fonctions de Bernstein. La nouvelle famille de bases obtenue
d\'epend d'un indice $\alpha \in \osegment{-\infty}{0}\cup  \osegment{1}{+\infty}$, et 
pour un degr\'e $k\in \NN^*$, est form\'ee de fonctions rationnelles dont le
num\'erateur et le d\'enominateur sont de degr\'e $k$. Ces bases de Bernstein rationnelles
v\'erifient toutes les propri\'et\'es classiques notamment la positivit\'e, la partition de l'unit\'e
et constituent de v\'eritables bases d'approximation de fonctions continues.

Les courbes de B\'ezier associ\'ees v\'erifient les propri\'et\'es classiques et on a obtenu 
les algorithmes d'\'evaluation comme celui de deCasteljau et de subdivision avec 
une complexit\'e \'equivalente.
Pour le m\^eme degr\'e $k$ et le m\^eme polygone de contr\^ole ces algorithmes convergent
vers la m\^eme courbe de B\'ezier que le cas standard.

Les bases de Bernstein polynomiales classiques se rev\`elent comme un cas asymptotique
de la nouvelle classe.

\end{abstract}

\noindent
\textbf{Mots cl\'es} : Fonctions de Bernstein rationnelles, Approximation de fonctions, 
Courbe de B\'ezier, Algorithme de deCasteljau.

\selectlanguage{english}

\begin{abstract}
In this paper, we construct a family of Bernstein functions using a class of 
rational parametrization.
The new family of rational Bernstein basis on an  
index $\alpha \in \osegment{-\infty}{0}\cup  \osegment{1}{+\infty}$, and
for a given  degree $k\in \NN^*$, these basis functions are  rational with 
a numerator and a denominator  are polynomials of degree $k$.
All of the classical properties 
as  positivity, partition of unity are hold for  these rational Bernstein  basis 
and they  constitute  approximation
basis functions for continuous functions spaces.

The  B\'ezier curves obtained verify  the classical properties and we have
the classical computational algorithms like the  deCasteljau Algorithm and the 
algorithm of subdivision with the similar accuracy.

Given a degree $k$ and a  control polygon points  all of these algorithms converge
to the same  B\'ezier curve as the classical case. That means the B\'ezier curve 
is independent of the index $\alpha$.

The classical polynomial Bernstein basis seems a asymptotic case of our new class
of rational Bernstein basis.

\end{abstract}


\tableofcontents


\section{Introduction}
We know that, for a B\'ezier curve $B$ of degree $n \in \NN^*$
in $\RR^d$ with $d \in \NN^*$ and $1\leq d \leq 3$,  the polynomial  Bernstein basis
$\suite{\Bezier{i}{n}{}}_{i=0}^{n}$ 
 can be define for $a<b \in \RR$  and a parametrization
$x \in [a \, , \, b]$, by :
\begin{equation} \label{BernsteinP}
\Bezier{i}{n}{}(x)=\left\lbrace
   {
   \begin{array}{ll}
   \displaystyle{{\left(\frac{b-x}{b-a} \right)}^{n}} 
   &\textrm{if }  i=0  \\
   C_{n}^{i}
   \displaystyle{{\left(\frac{x-a}{b-a} \right)}^{i}{\left(\frac{b-x}{b-a} \right)}^{n-i}} 
   &\textrm{if } 1 \leq i \leq n-1  \\
   \displaystyle{{\left(\frac{x-a}{b-a} \right)}^{n}} 
   &\textrm{if } i=n \\
   0 &\textrm{otherwise }
   \end{array}
   }
   \right.
\end{equation}
where 
$\displaystyle{C_{n}^{i} = \frac{n!}{i!(n-i)!} }$ and $0! =1$.

Let $\displaystyle{\suite{d_i}_{i=0}^{n}}$  be the control points of  
$B$, $d_i \in \RR^d$ for all $i$ then we have 
$$
B(x) = \displaystyle{\sum_{i=0}^{n} d_i\, \Bezier{i}{n}{} (x)} \, ,\,
\forall x \in [a \, , \, b]
$$

In the same way we define the rational Bernstein basis
$\suite{ R_i }_{i=0}^{n}$
  of degree $n \in \NN^*$ on $[a\, ,\,b]$ by 
$$
 R_i (x) =
\displaystyle{
\frac{\omega_i \Bezier{i}{n}{}(x)}{
\displaystyle{\sum_{j=0}^{n}\omega_j \Bezier{j}{n}{}(x)}
}
}
$$
where $\omega_i >0, \, \forall i=0, \ldots , n$.

 We can then define the rational B\'ezier curves  by substiting
the polynomial basis by the  rational one.

We observe the that, by putting $\displaystyle{w(x)=\frac{x-a}{b-a}}$ we have
 $\displaystyle{1-w(x)=\frac{b-x}{b-a}}$. 
 Therefore,
 $$
   \Bezier{i}{n}{}(x)=\left\lbrace
   {
   \begin{array}{ll}
  \displaystyle{C_{n}^{i}}
   \displaystyle{{\left( w(x) \right)}^{i}{\left( 1-w(x) \right)}^{n-i}} 
   &\textrm{if } 0 \leq i \leq n \textrm{ and } a  \leq x \leq b  \\
   0 &\textrm{otherwise }
   \end{array}
   }
   \right.
 $$

This function $w$ we defined is increasing on $[a \, , \, b]$ with
$w(a)=0$ and $w(b)=1$. 

The goal here is to retain these properties while requiring that $w$ is
homographic, to get a basis Bernstein naturally formed, 
of rational functions of degree $(k,k)$, 
ie a numerator of degree $k$ and a denominator of degree $k$.




\section{New class of rational B\'ezier like basis \label{SecNewClassBezier}}
\ifthenelse{\boolean{isthese}}{}{
\subsection{A class of  rational parametrizations}
We start with a lemma who fixed the foundation of our new class rational Bezier curves 
and can be expressed as follow :

\begin{lemme}
Let  $a,\, b \in \RR$ such that  $a < b$. There exists 
${\cal H}{\left( [a\, ,\, b]\right)}$ a family 
of the  homographic functions strictly increasing $f$
on $[a\, ,\, b]$ on $[a , b]$ such that $f(a)=0$ and $f(b)=1$.

More precisely, for all  $f\in {\cal H}{\left( [a\, ,\, b]\right)}$  there exists a unique 
$\alpha \in ]-\infty \, , \, 0[ \cup ]1  \, , \,  +\infty[ $ such that
$$
f(x) =\displaystyle{\frac{\alpha (x-a)}{x+(\alpha - 1)b-\alpha a}}, \quad 
\forall x \in [a\, ,\, b]
$$
\end{lemme}

\Preuve 
\emph{( Existence )} 

Since $f$ is homographic such that $f(a)=0$, then there exist $\alpha \neq 0$  and 
$c\in \RR \backslash \{ -a,\, -b\}$ 
such that for all   $x \in [a\, ,\, b]$   we have  :
$\displaystyle{f(x) = \frac{\alpha (x-a)}{x+c}}$. 
We have  $f(b)=1$ then  
$\displaystyle{1 = \frac{\alpha (b-a)}{b+c}}$. 
We get $c=(\alpha - 1)b-\alpha a$.
Since $c\notin \{ -a, \, -b\}$ then we have $\alpha \notin \{ 0, \, 1\}$. 
By strictly increasing of  $f$ we have $\alpha (\alpha -1) > 0$, and
 $\alpha \in ]-\infty \, , \, 0[ \cup ]1  \, , \,  +\infty[$.
 
 Therefore, we have 
 $$
{\cal H}{\left( [a\, ,\, b]\right)} =
\left\lbrace
\displaystyle{f_\alpha } \vert \,
\displaystyle{f_\alpha (x) }=
\displaystyle{\frac{\alpha (x-a)}{x+(\alpha - 1)b-\alpha a}}, \,
 \alpha \in ]-\infty \, , \, 0[ \cup ]1  \, , \,  +\infty[,
\, x \in  [a\, ,\, b]
\right\rbrace
$$

\emph{( Uniqueness)} 

Let  $\alpha, \, \beta \in ]-\infty \, , \, 0[ \cup ]1  \, , \,  +\infty[$ and 
$f_\alpha, \, f_\beta \in {\cal H}{\left( [a\, ,\, b]\right)}$ the associate homographic
functions
$$
\begin{array}{lcl}
f_\alpha = f_\beta
& \equivaut&
f_\alpha(x) =f_\beta(x) \quad \forall x\in  [a\, ,\, b]\\
&\implique&
\displaystyle{\frac{\alpha (x-a)}{x+(\alpha - 1)b-\alpha a}}=
\displaystyle{\frac{\beta (x-a)}{x+(\beta - 1)b-\beta a}}
\quad \forall x\in  [a\, ,\, b]\\
&\implique&
\displaystyle{(\alpha-\beta) (x-b)}=0
\quad \forall x\in  [a\, ,\, b]\\
&\implique&
\displaystyle{(\alpha-\beta) (x-b)}=0
\quad \forall x\in  [a\, ,\, b[\\
&\implique&
\displaystyle{\alpha=\beta}
\end{array}
$$

\begin{remarque}
 Let  $x \in [a\, ,\, b]$ and
  $\alpha \in ]-\infty \, , \, 0[ \cup ]1  \, , \,  +\infty[$.\\
  We have
 $D=x+(\alpha - 1)b-\alpha a \neq 0$.
 
 Ideed???, we observe that 
 $D=x- b+\alpha(b- a) =x- a+(\alpha - 1)(b- a)$,
  then  we have $ (\alpha-1)(b-a) \leq D \leq \alpha (b-a) $.  We can now write 
 $$
 \left\lbrace
 \begin{array}{ll}
 0 < \alpha(\alpha-1)(b-a) \leq \alpha D \leq \alpha^2 (b-a)& \textrm{ if } \alpha > 1\\
 0 < \alpha^2(b-a) \leq \alpha D \leq \alpha(\alpha-1) (b-a)& \textrm{ if } \alpha <0
 \end{array}
 \right.
 $$
Then we conclude that $D\neq 0 \, \forall x \in  [a\, ,\, b]$.
\end{remarque}

\begin{remarque} \label{RemValFalfa}
 Let  $\alpha \in ]-\infty \, , \, 0[ \cup ]1  \, , \,  +\infty[$ and $a<b$.\\
 Let $f_\alpha \in {\cal H}{\left( [a\, ,\, b]\right)}$  continuous and 
 increasing strictly on  $[a\, ,\, b]$ with $f_\alpha([a\, ,\, b])= [0\, ,\, 1]$.

 Moreover, for  $\lambda \in [0\, ,\, 1]$ and $x = a+\lambda(b-a) \in [a\, ,\, b]$ 
we have
$$
f_\alpha(x)=\displaystyle{\frac{\lambda\alpha}{\lambda+\alpha -1} \in  [0\, ,\, 1]}
$$

Thus obtaining the standard case as a asymptotic one :
$$
\displaystyle{\lim_{\module{\alpha} \to \infty}
f_\alpha(x)=\lambda=\frac{x-a}{b-a} }
$$
\end{remarque}

}
%
\subsection{A new class of rational Bernstein functions}
\begin{definition}
Let $a, b \in \RR$ such that $a<b$, 
 $\alpha \in ]-\infty \, , \, 0[ \cup ]1  \, , \,  +\infty[$,
 $f_\alpha \in {\cal H}{\left( [a\, ,\, b]\right)}$ and $n\in \NN^*$.

A  rational Bernstein  basis of  $\alpha$ index  and degree  $n$ on  $[a\, ,\, b]$,  is a family   
${\suite{\Bezier{i}{n}{\alpha}}}_{i=0}^{n}$ 
 of real functions defined for all  $x \in [a\, ,\, b]$  ;
\begin{equation}
\label{DefBase}
   \Bezier{i}{n}{\alpha}(x)=\left\lbrace
   {
   \begin{array}{ll}
   \displaystyle{{\left( 1-w(x) \right)}^{n}} 
   &\textrm{if }  i =0  \\
   \\
  \displaystyle{C_{n}^{i}}
   \displaystyle{{\left( w(x) \right)}^{i}{\left( 1-w(x) \right)}^{n-i}} 
   &\textrm{if } 1 \leq i \leq n-1  \\
   \\
   \displaystyle{{\left( w(x) \right)}^{n}} 
   &\textrm{if }  i = n  \\
   \\
   0 &\textrm{otherwise }
   \end{array}
   }
   \right.
\end{equation}
 where $\displaystyle{w(x)=f_{\alpha}(x)}$. 
 
 It was agreed that
 $\displaystyle{ \Bezier{0}{0}{\alpha}  (x)  =1, \forall x \in [a\, ,\, b] }$
 and  $[a\, ,\, b]$ the parametrization space ( or parameter space).
\end{definition}

\begin{definition}
Let $n, \, d \in \NN^*$ 
and $a, b \in \RR$  such that  $a<b$. \\
Let $\alpha \in ]-\infty \, , \, 0[ \cup ]1  \, , \,  +\infty[$
and $\suite{d_i}_{i=0}^{n} \subset \RR^d$.

A  rational B\'ezier curve of $\alpha$ index, degree  $n$ 
on $[a,b]$  and corresponding control polygon points $\suite{d_i}_{i=0}^{n}$,
the  function  $B_\alpha$  defined on $[a\, ,\, b]$ valued  in $\RR^d$  by 
$$
\displaystyle{
B_\alpha (x)=\sum_{i=0}^{n} d_i \,\Bezier{i}{n}{\alpha}(x) , \forall x \in [a\, ,\, b]
}
$$
where 
$\suite{\Bezier{i}{n}{\alpha}}_{i=0}^{n}$ is the  rational Bernstein  basis of  $\alpha$ index  and degree  $n$ as defined by the equation \ref{DefBase}
\end{definition}

\begin{remarque}
Observe that  $\suite{\Bezier{i}{n}{\infty}}_{i=0}^{n}$  is the classical polynomial Bernstein basis defined in \ref{BernsteinP}.
\end{remarque}

\begin{castest}
Before any deep analysis, we show throughout the  figures \ref{figBaseDegre1},
 \ref{figBaseDegre2},  \ref{figBaseDegre3},  \ref{figBaseDegre4} and
 \ref{figBaseDegre5}
 the  qualitative effect of the  index $\alpha$ on the behavior of the new 
 Bernstein basis functions.
 
\begin{figure}[h!]
\begin{center}
\includegraphics[width=11cm]{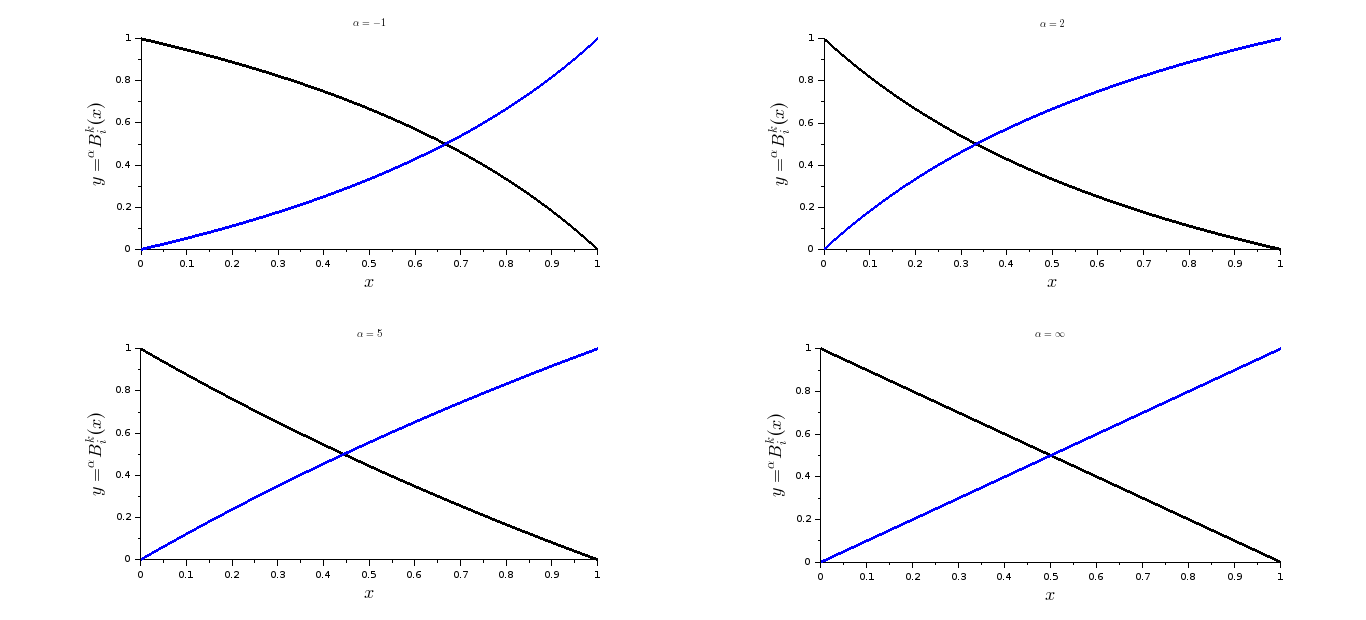}
\caption{Basis function $\displaystyle{\Bezier{i}{1}{\alpha}}$ for  $\alpha \in \lbrace-1, 2, 5, \infty \rbrace$}
\label{figBaseDegre1}
\end{center}
\end{figure}

\begin{figure}[h!]
\begin{center}
\includegraphics[width=11cm]{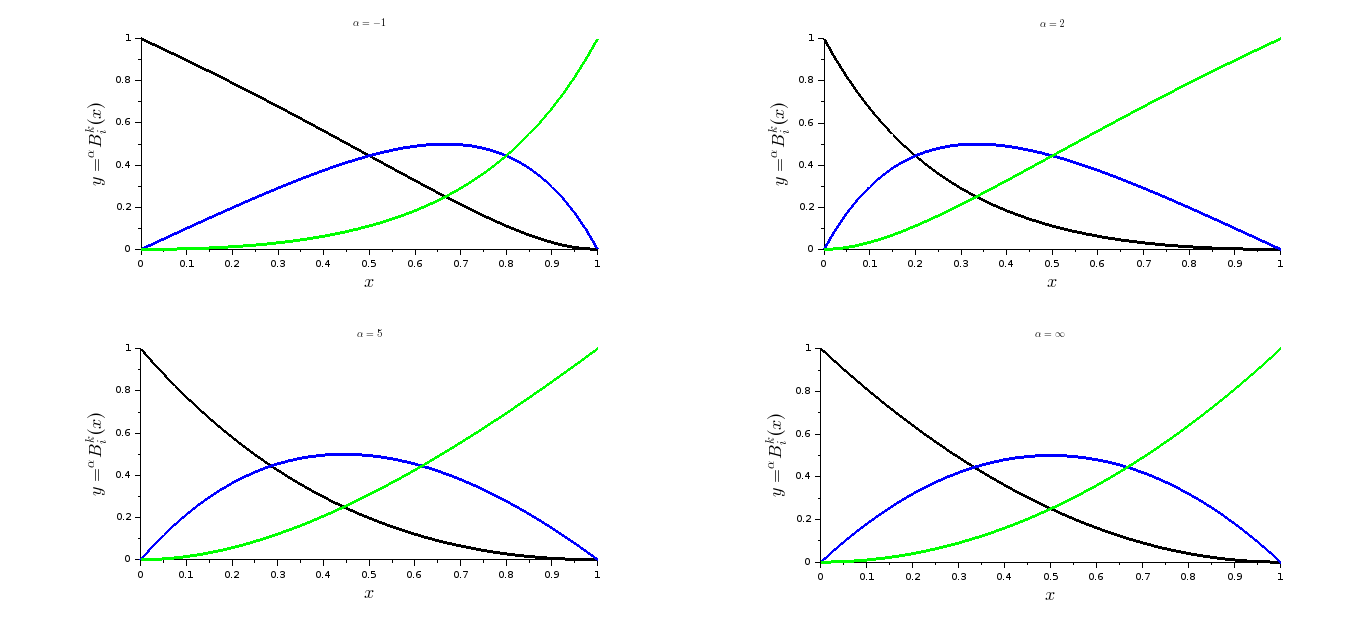}
\caption{Basis function $\displaystyle{\Bezier{i}{2}{\alpha}}$ for  $\alpha \in \lbrace-1, 2, 5, \infty \rbrace$}
\label{figBaseDegre2}
\end{center}
\end{figure}

\begin{figure}[h!]
\begin{center}
\includegraphics[width=11cm]{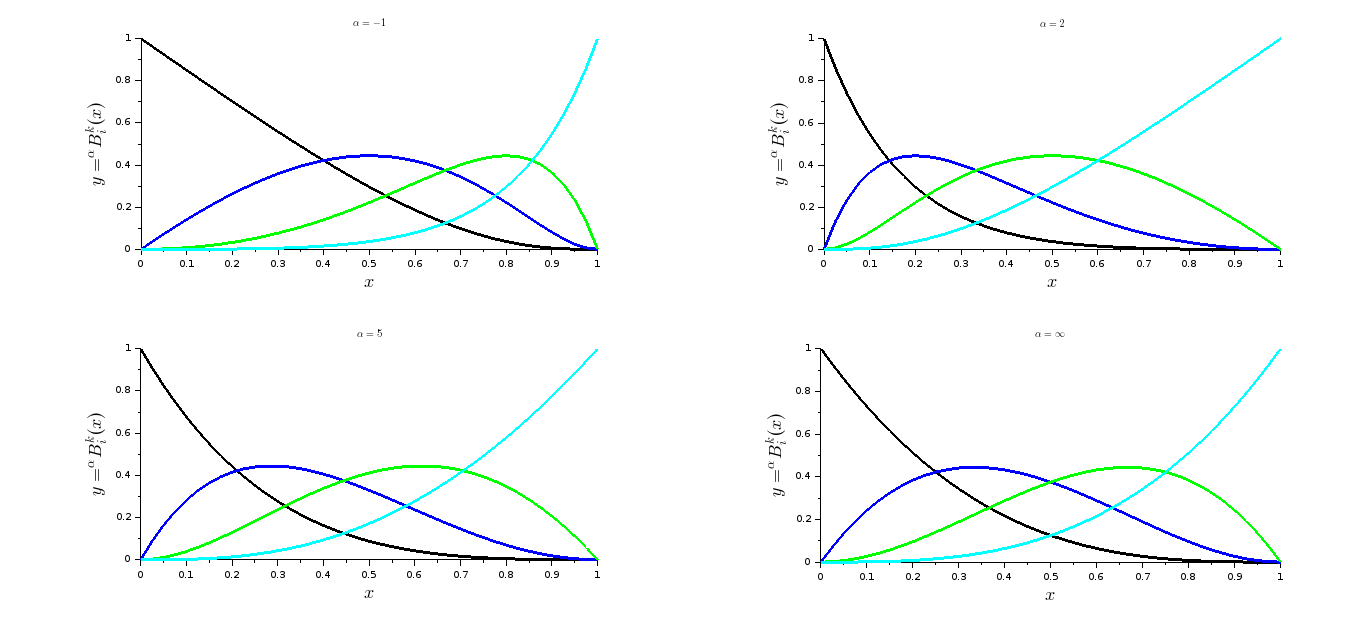}
\caption{Basis function $\displaystyle{\Bezier{i}{3}{\alpha}}$ for  $\alpha \in \lbrace-1, 2, 5, \infty \rbrace$}
\label{figBaseDegre3}
\end{center}
\end{figure}

\begin{figure}[h!]
\begin{center}
\includegraphics[width=11cm]{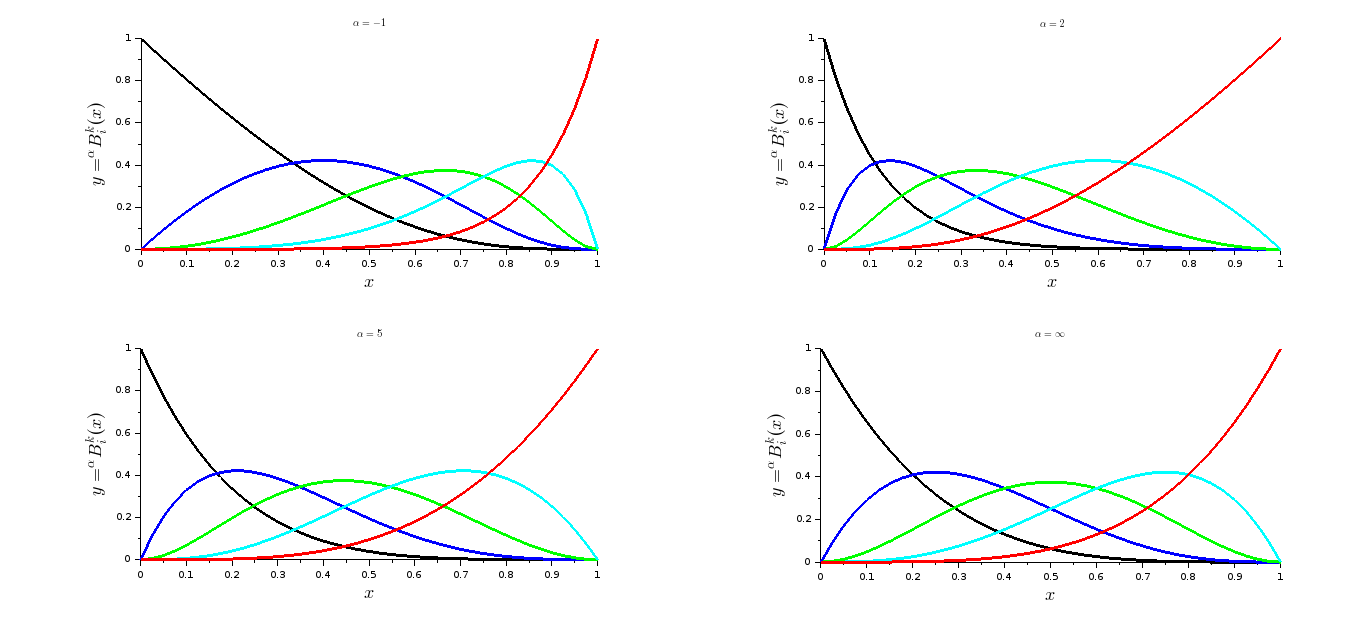}
\caption{Basis function $\displaystyle{\Bezier{i}{4}{\alpha}}$ for $\alpha \in \lbrace-1, 2, 5, \infty \rbrace$}
\label{figBaseDegre4}
\end{center}
\end{figure}

\begin{figure}[h!]
\begin{center}
\includegraphics[width=11cm]{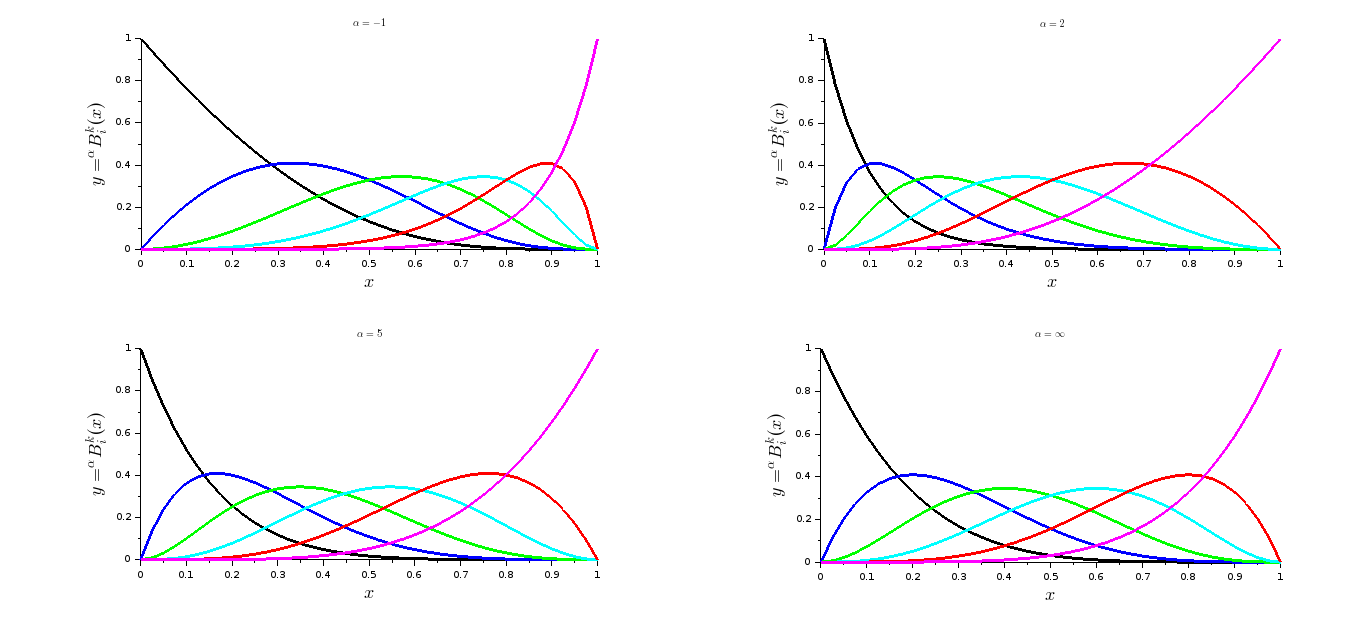}
\caption{Basis function $\displaystyle{\Bezier{i}{5}{\alpha}}$ for  $\alpha \in \lbrace-1, 2, 5, \infty \rbrace$}
\label{figBaseDegre5}
\end{center}
\end{figure}

A flash exploration of these figures shows that the  parameter $\alpha$
has a   notable effect on the functions of the  Bernstein basis for 
any  degree. We now refine this first reading by the check of the classical 
 properties of Bernstein functions.
\end{castest}

\subsection{Properties of the new  Bernstein basis}

\begin{proposition}
Let $n \in \NN^*$ 
and $a, b \in \RR$  such that  $a<b$. \\
Let $\alpha \in ]-\infty \, , \, 0[ \cup ]1  \, , \,  +\infty[$,
$f_\alpha \in {\cal H}{\left( [a\, ,\, b]\right)}$ and 
$\suite{\Bezier{i}{n}{\alpha}}_{i=0}^{n}$ the  rational Bernstein  basis of  $\alpha$ index  and degree  $n$  defined by the equation  \ref{DefBase}.

The following properties hold :
\begin{enumerate}
\item \emph{The positivity : }
For all $x \in [a\, ,\, b]$,  $\Bezier{i}{n}{\alpha} (x) \geq 0$
\item \emph{The partition of unity :}
For all  $x \in [a\, ,\, b]$,  
$\displaystyle{ \sum_{i=0}^{n}\Bezier{i}{n}{\alpha}  (x)  = 1}$
\item \emph{The symmetry :}
For all  $x \in [a\, ,\, b]$, for all  $i=0, \ldots , n$,\\
$\Bezier{i}{n}{\alpha}  (a+b-x)  = \Bezier{n-i}{n}{1-\alpha}  (x) $
\item \emph{The recursion  :}
For all  $x \in [a\, ,\, b]$, for all  $i=0, \ldots , n+1$,\\
$$
\displaystyle{ \Bezier{i}{n+1}{\alpha}  (x)  = 
w(x)\, \Bezier{i-1}{n}{\alpha}  (x) + 
(1-w(x))\, \Bezier{i}{n}{\alpha}  (x)
}
$$
where $\displaystyle{w(x)=f_\alpha (x)}$.


\end{enumerate}

\end{proposition}

\Preuve
Soit $\alpha \in ]-\infty \, , \, 0[ \cup ]1  \, , \,  +\infty[$
et $f_\alpha \in {\cal H}{\left( [a\, ,\, b]\right)}$.

\begin{enumerate}
\item (\emph{The positivity})

By remark ~\ref{RemValFalfa}, for all $x \in [a\, ,\, b]$
$\displaystyle{w(x)=f_\alpha (x) \in [0\, ,\, 1]}$. Then
$\displaystyle{1-w(x) \in [0\, ,\, 1]}$ so that 
$$
\displaystyle{
\Bezier{i}{n}{\alpha}  (x)= C_{n}^{i}w^{i}(x){\left( 1- w(x)\right)}^{n-i}  \geq 0
\quad\forall i=0, \ldots, n
}
$$

\item (\emph{The partition of unity})

Using the binomial formula of Newton's, we have 
$$
\begin{array}{rcl}
\displaystyle{\sum_{i=0}^{n} \Bezier{i}{n}{\alpha}(x)} 
&=&
\displaystyle{\sum_{i=0}^{n}  C_{n}^{i}w^{i}(x){\left( 1- w(x)\right)}^{n-i}} \\
&=&\displaystyle{{\left(w(x)+ (1- w(x))\right)}^{n}=1}\\
\end{array}
$$
\item (\emph{The symmetry})

Let $i=0, \ldots, n$ and $x\in [a\, , \, b]$. We have $y=a+b-x\in [a\, , \, b]$ and
$$
\left\lbrace
\begin{array}{rcl}
\displaystyle{w(y)=f_\alpha (a+b-x)}&=&
\displaystyle{\frac{\alpha(b-x)}{-x+\alpha b +(1-\alpha)a}}\\
1-w(y)&=&
\displaystyle{\frac{(1-\alpha)(x-a)}{x-\alpha b -(1-\alpha)a}}\\
&=&\displaystyle{f_{1-\alpha}(x)}\\
\end{array}
\right.
$$
Then
$$
\begin{array}{rcl}
\displaystyle{\Bezier{n-i}{n}{\alpha}(a+b-x)} 
&=&
\displaystyle{\Bezier{n-i}{n}{\alpha}(y)} \\
&=&
\displaystyle{  C_{n}^{n-i}w^{n-i}(y){\left( 1- w(y)\right)}^{i}} \\
&=&
\displaystyle{  C_{n}^{n-i}{\left( 1- f_{1-\alpha}(x)\right)}^{n-i}
{\left(f_{1-\alpha}(x)\right)}^{i}} \\
&=&
\displaystyle{  C_{n}^{i}{\left(f_{1-\alpha}(x)\right)}^{i}
{\left( 1- f_{1-\alpha}(x)\right)}^{n-i}}\\ 
&=&\displaystyle{\Bezier{i}{n}{1-\alpha}(x)} \\
\end{array}
$$
\item (\emph{The recursion})
$$
\begin{array}{rcl}
\displaystyle{\Bezier{i+1}{n+1}{\alpha}(x)} 
&=&
\displaystyle{  C_{n+1}^{i+1}w^{i+1}(x){\left( 1- w(x)\right)}^{n-i}} \\
&=&
\displaystyle{ {\left( C_{n}^{i}+ C_{n}^{i+1}\right)} w^{i+1}(x){\left( 1- w(x)\right)}^{n-i}} \\
&=&
\displaystyle{  C_{n}^{i} w^{i+1}(x){\left( 1- w(x)\right)}^{n-i}}+
\displaystyle{ C_{n}^{i+1} w^{i+1}(x){\left( 1- w(x)\right)}^{n-i}} \\
&=&
\displaystyle{ w(x) C_{n}^{i} w^{i}(x){\left( 1- w(x)\right)}^{n-i}}+
\displaystyle{{\left( 1- w(x)\right)} C_{n}^{i+1} w^{i+1}(x){\left( 1- w(x)\right)}^{n-i-1}} \\
&=&
\displaystyle{w(x)\, \Bezier{i}{n}{\alpha}(x)} +
\displaystyle{{\left( 1- w(x)\right)}\, \Bezier{i+1}{n}{\alpha}(x)} \\
\end{array}
$$

\end{enumerate}

\begin{remarque}
For all $n \in \NN^*$, 
 $a, b \in \RR$ such that $a<b$ and\\
 $\alpha \in ]-\infty \, , \, 0[ \cup ]1  \, , \,  +\infty[$.
 $\suite{\Bezier{i}{n}{\alpha}}_{i=0}^{n}$, the rational Bernstein basis of
 $\alpha$ index and degree   $n$ on  the parameter space $[a\, , \, b]$.
satisfies :
\begin{equation}
\label{EndValBase}
\left\lbrace
\begin{array}{l}
\displaystyle{ \Bezier{0}{n}{\alpha}  (a)  = \Bezier{n}{n}{\alpha}  (b)  =1 }\\
\displaystyle{ \Bezier{i}{n}{\alpha}  (a)  =0 } \quad \forall i=1, \ldots, n\\
\displaystyle{ \Bezier{i}{n}{\alpha}  (b)  =0 } \quad \forall i=0, \ldots, n-1\\
\displaystyle{ \Bezier{i}{n}{\alpha}  (x)  >0 } \quad \forall i=0, \ldots, n,
 \, \forall x \in ]a\, , \, b[\\
\end{array}
\right.
\end{equation}
\end{remarque}

\begin{proposition}
\label{DegrePlus}
Let $n \in \NN^*$ and
 $a, b \in \RR$ such that $a<b$.\\
Let $\alpha \in ]-\infty \, , \, 0[ \cup ]1  \, , \,  +\infty[$.
and $f_\alpha \in {\cal H}{\left( [a\, ,\, b]\right)}$.

 Consider the rational Bernstein basis 
$\suite{\Bezier{i}{n}{\alpha}}_{i=0}^{n}$ and 
$\suite{\Bezier{i}{n+1}{\alpha}}_{i=0}^{n+1}$ 
of  $\alpha$ index, of degrees  $n$ and $n+1$ with parameters in $[a\, , \, b]$.

For all $x \in [a\, , \, b]$ and $\displaystyle{w(x)=f_\alpha (x)}$,
we have :
\begin{enumerate}
\item
$\displaystyle{
{\left(1-w(x)\right)\, }\Bezier{i}{n}{\alpha}(x) = 
\frac{n+1-i}{n+1}\Bezier{i}{n+1}{\alpha}
}$, $\forall i =0, \ldots , n$ 
\item
$\displaystyle{
{w(x)\,}\Bezier{i}{n}{\alpha}(x) = 
\frac{i+1}{n+1}\Bezier{i+1}{n+1}{\alpha}
}$, $\forall i =0, \ldots , n$ 
\end{enumerate}

\end{proposition}

\Preuve
By a direct computation.

\begin{proposition}
Let $n \in \NN^*$ and
 $a, b \in \RR$ such that $a<b$.\\
Let $\alpha \in ]-\infty \, , \, 0[ \cup ]1  \, , \,  +\infty[$.
and $f_\alpha \in {\cal H}{\left( [a\, ,\, b]\right)}$.

The rational Bernstein basis
$\suite{\Bezier{i}{n}{\alpha}}_{i=0}^{n}$ of $\alpha$ index and degree $n$ 
on  the parameter space $[a\, , \, b]$,
is formed of rational functions of degree  $\couple{n}{n}$ and 
 infinitely differentiable on $[a,b]$. Moreover, we have :
\begin{enumerate}
\item \emph{First order  derivative : }
For all $x \in [a\, ,\, b]$ and $i=0, \ldots , n$ we have
\begin{equation}
\label{DerivBase}
 \frac{d}{dx} \Bezier{i}{n}{\alpha} (x) =
 \left\lbrace
 \begin{array}{ll}
\displaystyle{-n\frac{d w}{dx} {\left( 1 - w(x)\right)}^{n-1}}& i=0\\
\\
\displaystyle{
{\left( i- n w(x)\right)}\frac{d w}{dx} \times 
}&\\
\displaystyle{{C_{n}^{i}}w^{i-1}(x) {\left( 1 - w(x)\right)}^{n-i-1}}
& 1\leq i \leq n-1\\
\\
\displaystyle{n \frac{d w}{dx}w^{n-1}(x)} &  i = n\\
\end{array}
\right.
\end{equation}
with
$\displaystyle{w(x)=f_\alpha(x)=\frac{\alpha(x-a)}{x+(\alpha-1)b-\alpha a}}$
and
$\displaystyle{\frac{dw}{dx}(x)=
\frac{\alpha(\alpha-1)(b-a)}{\left(x+(\alpha-1)b-\alpha a \right)^2}
}$
\item \emph{Seconde order DerivBase : }
For all $x \in [a\, ,\, b]$ and $i=0, \ldots , n$
$$
 \frac{d^2}{dx^2} \Bezier{i}{n}{\alpha} (x) =
 \left\lbrace
 \begin{array}{ll}
\displaystyle{n{\left( 1 - w(x)\right)}^{n-2}
{\left[-\frac{d^2 w}{dx^2} ( 1 - w(x)) 
+(n-1)\left(\frac{d w}{dx} \right)^2\right]}
}& i=0\\
\displaystyle{n{\left( 1 - w(x)\right)}^{n-3}
{\left[\frac{d^2 w}{dx^2} ( 1 - w(x)) ( 1 -n w(x)) 
\right.}}
\\
\displaystyle{
{\left.
-(n-1)\left(\frac{d w}{dx} \right)^2( 2 -n w(x))\right]}
}& i=1\\
\displaystyle{
{C_{n}^{i}}w^{i-2}(x) {\left( 1 - w(x)\right)}^{n-i-2}\times
}&\\
\displaystyle{\left[ {\left( i - n w(x)\right)}w(x){\left( 1 - w(x)\right)}\frac{d^2 w}{dx^2}
\right.
}\\
\displaystyle{
\left.
+\left\lbrace
(n-1){\left(n w(x)-2i\right)}w(x)+i(i-1)
\right\rbrace
{\left( \frac{d w}{d x}\right)}^2
\right]
}
& 2\leq i \leq n-2\\
\end{array}
\right.
$$
and
$$
 \frac{d^2}{dx^2} \Bezier{i}{n}{\alpha} (x) =
 \left\lbrace
 \begin{array}{ll}
\displaystyle{n w^{n-3}(x)
{\left[\frac{d^2 w}{dx^2} w(x) (n-1-nw(x))
\right.}}
\\
\displaystyle{
{\left.
+(n-1)\left(\frac{d w}{dx} \right)^2 (n-2-nw(x))\right]}
}&  i = n-1\\
\displaystyle{n w^{n-2}(x)
{\left[\frac{d^2 w}{dx^2} w(x) 
+(n-1)\left(\frac{d w}{dx} \right)^2\right]}
}&  i = n\\
\end{array}
\right.
$$

where 
$\displaystyle{
\frac{d^2 w}{dx^2} =
\frac{-2\alpha(\alpha-1)(b-a)}{\left(x+(\alpha-1)b-\alpha a \right)^3}
}$
\end{enumerate}

\end{proposition}

\Preuve
$\displaystyle{\Bezier{i}{n}{\alpha}}$  is a polynomial of 
$f_\alpha$. Since
$\displaystyle{ f_\alpha \in {\cal C}^\infty {\left( [a\, , \, b] \right)}}$, 
it is the same for  $\displaystyle{\Bezier{i}{n}{\alpha}}$.

We complete the proof by simple computation.

\begin{remarque}
For all $n \in \NN^*$ and 
 $a, b \in \RR$ such that $a<b$ and\\
 $\alpha \in ]-\infty \, , \, 0[ \cup ]1  \, , \,  +\infty[$,
 using the relation \ref{DerivBase}, we observe that the derivatives of
 elements of the rational  Bernstein basis of $\alpha$ index and degree $n$ 
 on  the parameter space $[a\, , \, b]$,
$\suite{\Bezier{i}{n}{\alpha}}_{i=0}^{n}$ satisty the following relations  :
\begin{equation}
\label{EndDerivBase}
\left\lbrace
\begin{array}{l}
\displaystyle{\frac{d}{dx}\Bezier{0}{n}{\alpha}  (a)  = -n \frac{\alpha}{(\alpha-1)(b-a)} }
\textrm{ and }
\displaystyle{\frac{d}{dx} \Bezier{1}{n}{\alpha}  (a)  = n \frac{\alpha}{(\alpha-1)(b-a)} }\\
\\
\displaystyle{\frac{d}{dx} \Bezier{n-1}{n}{\alpha}  (b)  =n  \frac{\alpha-1}{\alpha(b-a)} }
\textrm{ and }
\displaystyle{\frac{d}{dx} \Bezier{n}{n}{\alpha}  (b)  =-n  \frac{\alpha-1}{\alpha(b-a)} }\\
\\
\displaystyle{\frac{d}{dx} \Bezier{i}{n}{\alpha}  (a)  =0 } \quad \forall i=2, \ldots, n
\textrm{ and }
\displaystyle{\frac{d}{dx} \Bezier{i}{n}{\alpha}  (b)  =0 } \quad \forall i=0, \ldots, n-2\\
\end{array}
\right.
\end{equation}
\end{remarque}

\begin{proposition}
Let $n \in \NN^*$ and  $a, b \in \RR$ such that $a<b$ and \\
 $\alpha \in ]-\infty \, , \, 0[ \cup ]1  \, , \,  +\infty[$.
Let 
$\suite{\Bezier{i}{n}{\alpha}}_{i=0}^{n}$ be the rational Bernstein 
basis of $\alpha$ index and  degree $n$ on the parameter space $[a\, , \, b]$,
\begin{enumerate}
\item \emph{ Uniqueness of the maximun  : }
For $i=0, \ldots , n$,  $\Bezier{i}{n}{\alpha}$  has a unique maximun
 in $x_i \in [a\, ,\, b]$, such that $\displaystyle{w(x_i)=\frac{i}{n}}$. 
The value of this maximun   is independant of $\alpha$.

More precisely, we have
$$
x_i=\displaystyle{a+ \frac{i(\alpha -1)}{n\alpha-i}(b-a)}
\textrm{ and }
\Bezier{i}{n}{\alpha}(x_i)= 
\displaystyle{C_{n}^{i}\frac{i^i{(n-i)}^{n-i}}{n^n}}
$$
\item \emph{Maximun symmetrical property  : }
For $i=0, \ldots , n$,  $\Bezier{i}{n}{\alpha}$ ,  $\Bezier{n-i}{n}{\alpha}$  and
 $\Bezier{i}{n}{1-\alpha}$  have the same maximun.
\end{enumerate}

\end{proposition}

\Preuve

\begin{enumerate}
\item \emph{Existence and uniqueness of the  maximum}
\begin{itemize}
\item 
$\Bezier{0}{n}{\alpha}$ is decreasing strictly on $[a\, , \, b]$, then it has a 
maximun in $a$.
\item
$\Bezier{n}{n}{\alpha}$ is increasing strictly on $[a\, , \, b]$ , then it has a 
maximun in $b$.
\item 
For $i=1, \ldots, n-1$ the sign of 
$\displaystyle{\frac{d}{dx}\Bezier{i}{n}{\alpha}}$ and  
$ i- n w(x)$ are the same on $[a\, , \, b]$. 
This is the sign  same of $\displaystyle{ i- n f_\alpha(x)}$.

Since $f_\alpha$ is increasing strictly on  $[a\, , \, b]$ then there exist a unique
$x_i \in [a\, , \, b]$ such that $\displaystyle{ f_\alpha (x_i)=\frac{i}{n}\in [0\, , \, 1] }$.
We have
 $$
 \left\lbrace
 \begin{array}{ll}
  i- n w(x) >0& \forall x < x_i \\
  i- n w(x) <0& \forall x > x_i \\
 \end{array} 
 \right.
 $$
 Thus $\displaystyle{\Bezier{i}{n}{\alpha}}$ has a maximun in  $x_i$ satisfying
 $\displaystyle{ w(x_i)=\frac{i}{n} }$ and
 $\displaystyle{\Bezier{i}{n}{\alpha} (x_i)}$ is  independant of $\alpha$.
  and it can be obtained by simple computation
 
\end{itemize}
\item \emph{Maximun symmetrical property  : }

By simple computation, we have 
$$
\displaystyle{\max_{x\in [a\, , \, b]}\Bezier{i}{n}{\alpha} (x)} =
\displaystyle{\max_{x\in [a\, , \, b]}\Bezier{n-i}{n}{\alpha} (x)} =
\displaystyle{\max_{x\in [a\, , \, b]}\Bezier{i}{n}{1-\alpha} (x)} 
\quad \forall i=0, \ldots, n
$$
\end{enumerate}

\begin{proposition}
Let $n \in \NN^*$ and
 $a, b \in \RR$ such that $a<b$.\\
Let $\alpha \in ]-\infty \, , \, 0[ \cup ]1  \, , \,  +\infty[$.
and $f_\alpha \in {\cal H}{\left( [a\, ,\, b]\right)}$.

The rational Bernstein 
basis of $\alpha$ index and  degree $n$ on $[a\, , \, b]$,
$\suite{\Bezier{i}{n}{\alpha}}_{i=0}^{n}$  is a linear independent family. 

\end{proposition}

\Preuve
We proceed by induction  on  $n$.

Let $x \in [a\, , \, b]$ and $\displaystyle{w(x)=f_\alpha (x)}$

For $n=1$, if 
$\displaystyle{\sum_{i=0}^{n} \lambda_i \Bezier{i}{n}{\alpha} =0 }$ then for all   $x \in [a\, , \, b]$ we have
$\displaystyle{ \lambda_{0}{\left(1-w(x) \right)}+\lambda_{1}{w(x)} =0 }$.
In particular  :
$$
\left\lbrace
\begin{array}{l}
\displaystyle{ \lambda_{0}=\lambda_{0}{\left(1-w(a) \right)}+\lambda_{1}{w(a)} =0 }\\
\displaystyle{ \lambda_{1}=\lambda_{0}{\left(1-w(b) \right)}+\lambda_{1}{w(b)} =0 }\\
\end{array}
\right.
$$
Therefore 
$\suite{\Bezier{i}{1}{\alpha}}_{i=0}^{1}$ is a linear  independent family

Assumes that the property holds up to an order  $n\in \NN*$, and 
 show that $\suite{\Bezier{i}{n+1}{\alpha}}_{i=0}^{n+1}$  
 is a linear  independent family.

If $\displaystyle{\sum_{i=0}^{n+1} \lambda_i \Bezier{i}{n+1}{\alpha} =0 }$
then for all  $x \in ]a\, , \, b[$ we have 
$$
\begin{array}{rcl}
0&=& 
\displaystyle{\sum_{i=0}^{n+1} \lambda_i \Bezier{i}{n+1}{\alpha}(x) }\\
&=&
\displaystyle{\sum_{i=0}^{n+1} \lambda_i 
{\left( 
w(x)\Bezier{i-1}{n}{\alpha}(x) +
{\left(1- w(x)\right)}\Bezier{i}{n}{\alpha}(x)
\right)}
}\\
&=&
\displaystyle{\sum_{i=0}^{n}
{\left( 
w(x) \lambda_{i+1}+
{\left(1- w(x)\right)}\lambda_{i}
\right)}
\Bezier{i}{n}{\alpha}(x)
}\\

\end{array}
$$
Using the assumption :
$\suite{\Bezier{i}{n}{\alpha}}_{i=0}^{n}$ is a linear independent system
 we have 
\begin{equation}\label{eqLibre1}
w(x) \lambda_{i+1}+
{\left(1- w(x)\right)}\lambda_{i}=0,
\quad \forall i=0, \ldots, n, \, \forall x \in ]a\, , \, b[
\end{equation}
Otherwise
\begin{equation}\label{eqLibre2}
\displaystyle{ 
\lambda_{n+1}=
\sum_{i=0}^{n+1} \lambda_i \Bezier{i}{n+1}{\alpha}(b)=0 
}
\end{equation}

The linear system from the  equation ~\ref{eqLibre1} completed 
by equation~\ref{eqLibre2} is inversible triangular upper  since the terms 
of the diagonal are nonzero. 
We deduce that
$\displaystyle{\lambda_i =0, \quad \forall i=0, \ldots, n+1}$. 
Then we can conclude that 
$\suite{\Bezier{i}{n+1}{\alpha}}_{i=0}^{n+1}$  is a linear independent system.

\begin{remarque}
Since $\suite{\Bezier{i}{n}{\alpha}}_{i=0}^{n}$ is a linear independent system 
then there is a basis of an approximation space of continuous real functions.
\end{remarque}

\section{Properties of the new class of B\'ezier curves
\label{SecNewCourbBezier}}

Let $n \in \NN^*$ and
 $a, b \in \RR$ such that $a<b$.\\
Let $\alpha \in ]-\infty \, , \, 0[ \cup ]1  \, , \,  +\infty[$.
and $f_\alpha \in {\cal H}{\left( [a\, ,\, b]\right)}$.

Let $\suite{\Bezier{i}{n}{\alpha}}_{i=0}^{n}$ 
be the rational Bernstein basis of $\alpha$ index and  degree $n$ on $[a\, , \, b]$.

Consider the rational B\'ezier curve  $B_\alpha$ of $\alpha$ index and degree $n$
with control polygon points $\displaystyle{\suite{d_i}_{i=0}^{n}  \subset \RR^d}$ 
 defined for all $x \in [a\, , \, b]$ by 
$\displaystyle{B_\alpha(x) = \sum_{i=0}^{n}d_i \Bezier{i}{n}{\alpha}(x) }$

\subsection{Geometric properties}
The curves of this new class check the properties of the classical  B\'ezier curves .
The proposition below lists the most important of these properties.
\begin{proposition}
Let $B_\alpha$ the rational B\'ezier curve  of $\alpha$ index and  degree $n$
 with control polygon points  
$\displaystyle{\suite{d_i}_{i=0}^{n}}$  on  a parameter space $[a ,b]$.
 
The following properties hold :
 \begin{enumerate}
\item \emph{Extremities interpolation properties :} The B\'ezier curve  $B_\alpha$ 
interpol the extremities   points of its control polygon; 
it means that
$B_\alpha(a)=d_0$ and $B_\alpha(b)=d_n$

\item \emph{Extremities tangents properties :}
The B\'ezier curve  $B_\alpha$ is tangent to its control polygon 
  to the extremities. More precisely, we have 
$$
\left\lbrace
\begin{array}{l}
\displaystyle{
\frac{d B_\alpha}{dx}(a)=n
\frac{\alpha}{(\alpha-1)(b-a)}{\left( d_1- d_0\right)}
}
\\
\\
\displaystyle{
\frac{d B_\alpha}{dx}(b)=n
\frac{(\alpha-1)}{\alpha(b-a)}{\left( d_{n} - d_{n-1}\right)}
}
\end{array}
\right.
$$ 

\item \emph{Convex hull property :}
$B_\alpha$ is in the convex hull of its control points  $\displaystyle{\suite{d_i}_{i=0}^{n}}$. In the other way,
for all  $x \in [a\, , \, b]$, there exists
$\suite{\lambda_i}_{i=0}^{n} \subset \RR_+$ such that
$
B_\alpha(x)=
\displaystyle{\sum_{i=0}^{n}{\lambda_i d_i}}
$ 
with
$\displaystyle{\sum_{i=0}^{n}{\lambda_i}=1}$

\item \emph{Invariance property of affine transformation :}
For all affine transformation  $T$ in $\RR^d$, we have 
$$
T\left(B_\alpha(x)\right) =
\displaystyle{
\sum_{i=0}^{n}T(d_i)\Bezier{i}{n}{\alpha}(x)
}
$$

\end{enumerate}

\end{proposition}

\Preuve
Consider   the rational B\'ezier curve $B_\alpha$ of  $\alpha$  index
and degree $n$ with control polygon points 
$\displaystyle{\suite{d_i}_{i=0}^{n}}$ on the parameter space $[a ,b]$.
\begin{enumerate}
\item
Using the  remark (\ref{EndValBase}), we have
$$
\begin{array}{rcl}
B_\alpha(a)&=&\displaystyle{\sum_{i=0}^{n} d_{i}\Bezier{i}{n}{\alpha}(a)}\\
&=&\displaystyle{d_{0}\Bezier{0}{n}{\alpha}(a)=d_{0}}\\
\end{array}
$$
and
$$
\begin{array}{rcl}
B_\alpha(b)&=&\displaystyle{\sum_{i=0}^{n} d_{i}\Bezier{i}{n}{\alpha}(b)}\\
&=&\displaystyle{d_{n}\Bezier{n}{n}{\alpha}(b) =d_{n}}\\
\end{array}
$$

\item
Using the  remark  (\ref{EndDerivBase}), we have
$$
\begin{array}{rcl}
\displaystyle{\frac{d}{dx} B_\alpha(a)}&=&
\displaystyle{\sum_{i=0}^{n} d_{i}\frac{d}{dx}\, \Bezier{i}{n}{\alpha}(a)}\\
&=&\displaystyle{d_{0}\frac{d}{dx}\, \Bezier{0}{n}{\alpha}(a)+
d_{1}\frac{d}{dx}\, \Bezier{1}{n}{\alpha}(a)}\\
&=&\displaystyle{\left( d_{1}- d_{0}\right) \frac{d}{dx}\, \Bezier{1}{n}{\alpha}(a)}\\
\end{array}
$$
and
$$
\begin{array}{rcl}
\displaystyle{\frac{d}{dx} B_\alpha(b)}&=&
\displaystyle{\sum_{i=0}^{n} d_{i}\frac{d}{dx}\, \Bezier{i}{n}{\alpha}(b)}\\
&=&\displaystyle{d_{n-1}\frac{d}{dx}\, \Bezier{n-1}{n}{\alpha}(b)+
d_{n}\frac{d}{dx}\, \Bezier{n}{n}{\alpha}(b)}\\
&=&\displaystyle{\left( d_{n}- d_{n-1}\right) \frac{d}{dx}\, \Bezier{n}{n}{\alpha}(b)}\\
\end{array}
$$

\item
For all  $x\in [a\, , \, b]$, we have
$$
\left\lbrace
\begin{array}{rcl}
\displaystyle{B_\alpha (x)}
&=&
\displaystyle{\sum_{i=0}^{n} d_{i}  \Bezier{i}{n}{\alpha}(x)}\\
&=&
\displaystyle{\sum_{i=0}^{n}\lambda_i d_{i} }\\
\textrm{where}&&\\
\displaystyle{\lambda_i}
&=&
\displaystyle{ \Bezier{i}{n}{\alpha}(x)\in \RR_+  \, \forall i}\\
\end{array}
\right.
$$
Using a unity partition property we obtain
$
\displaystyle{\sum_{i=0}^{n}\lambda_i  
=\sum_{i=0}^{n}  \Bezier{i}{n}{\alpha}(x) =1 }
$. Then $B_\alpha(x)$  is in the convex hull of its control polygon points.
$\displaystyle{\suite{d_i}_{i=0}^{n}}$

\item
Let  $T$ be an affine transformation in $\RR^d$. There is a square matrix  $M$
of order $d$ and a point $C\in \RR^d$ such that for all   $X\in \RR^d$, we have
$\displaystyle{T(X)=M\,X+C}$.
Let $x\in [a\, , \, b]$. Since $\displaystyle{B_\alpha (x) \in \RR^d}$ we have
$$
\left\lbrace
\begin{array}{rcl}
\displaystyle{T{\left( B_\alpha (x)\right)}}
&=&
\displaystyle{T{\left( \sum_{i=0}^{n} d_{i}  \Bezier{i}{n}{\alpha}(x)\right)}}\\
&=&
\displaystyle{M{\left( \sum_{i=0}^{n} d_{i}  \Bezier{i}{n}{\alpha}(x)\right)}+C}\\
&=&
\displaystyle{\sum_{i=0}^{n}M{\left(  d_{i}  \Bezier{i}{n}{\alpha}(x)\right)}}+
\displaystyle{{\left( \sum_{i=0}^{n}\Bezier{i}{n}{\alpha}(x)\right)}C}\\
&=&
\displaystyle{\sum_{i=0}^{n}{\left( M d_{i}  \Bezier{i}{n}{\alpha}(x)\right)}}+
\displaystyle{ \sum_{i=0}^{n}{\left(C\Bezier{i}{n}{\alpha}(x)\right)}}\\
&=&
\displaystyle{ \sum_{i=0}^{n}{\left(M d_{i} +C\right)}\Bezier{i}{n}{\alpha}(x)}=
\displaystyle{ \sum_{i=0}^{n}T{\left(d_{i} \right)}\Bezier{i}{n}{\alpha}(x)}\\
\end{array}
\right.
$$
This proves the property.
\end{enumerate}

\begin{proposition}[Degree elevation algorithm]
Given  the rational B\'ezier curve  $B_\alpha$ of $\alpha$ index and
 degree $n$ with  control polygon points 
$\displaystyle{\suite{d_i}_{i=0}^{n}\subset \RR^d }$ on the parameter space $[a ,b]$,
then there exists  a rational B\'ezier curve $C_\alpha$ of $\alpha$ index and
 degree $n+1$ with  control polygon points 
$\displaystyle{\suite{\hat{d}_i}_{i=0}^{n+1}\subset \RR^d }$ 
on the parameter space $[a ,b]$ identical to 
$B_\alpha$. 

More precisely we have : 
$
B_\alpha (x) =\displaystyle{
\sum_{i=0}^{n+1}\hat{d}_i \Bezier{i}{n+1}{\alpha}(x)
}
$
with
$$
\left\lbrace
\begin{array}{ll}
\displaystyle{\hat{d}_{0}=d_{0} }&\\
\displaystyle{\hat{d}_{i}={\frac{i}{n+1}}d_{i-1} +{\left(1-\frac{i}{n+1}\right)}d_{i} }
& 1\leq i \leq n\\
\displaystyle{\hat{d}_{n+1}=d_{n}} &\\
\end{array}
\right.
$$

\end{proposition}

\Preuve
For all $\alpha \in ]-\infty \, , \, 0[ \cup ]1  \, , \,  +\infty[$
and $f_\alpha \in {\cal H}{\left( [a\, ,\, b]\right)}$, 
 putting $\displaystyle{w(x)=f_\alpha (x)}$ for all  $x\in [a\, , \, b]$.
Using the  proposition~\ref{DegrePlus}, we have :
$$
\begin{array}{rcl}
\displaystyle{ \Bezier{i}{n}{\alpha}(x)} 
&=&
{\left( 1 - w(x)\right)}\displaystyle{ \Bezier{i}{n}{\alpha}(x)}+
{w(x)}\displaystyle{ \Bezier{i}{n}{\alpha}(x)}
\quad \forall i=0, \ldots, n\\
&=&
\displaystyle{\frac{n+1-i}{n+1} \Bezier{i}{n+1}{\alpha}(x)}+
\displaystyle{\frac{i+1}{n+1} \Bezier{i+1}{n+1}{\alpha}(x)}
\quad \forall i=0, \ldots, n\\
\end{array}
$$
Then
$$
\begin{array}{rcl}
B_\alpha (x) &=&
\displaystyle{
\sum_{i=0}^{n}d_{i} \, \Bezier{i}{n}{\alpha}(x)
}\\
&=&
\displaystyle{
\sum_{i=0}^{n}d_{i} \,
\left[
{\frac{n+1-i}{n+1}}\, \Bezier{i}{n+1}{\alpha}(x) +
{\frac{i+1}{n+1}}\, \Bezier{i+1}{n+1}{\alpha}(x)
\right]
}\\
&=&
\displaystyle{
\sum_{i=0}^{n}d_{i} \,
{\frac{n+1-i}{n+1}}\, \Bezier{i}{n+1}{\alpha}(x)
}+
\displaystyle{
\sum_{i=0}^{n}d_{i} \,
{\frac{i+1}{n+1}}\, \Bezier{i+1}{n+1}{\alpha}(x)
}\\
&=&
\displaystyle{
\sum_{i=0}^{n}d_{i} \,
{\frac{n+1-i}{n+1}}\, \Bezier{i}{n+1}{\alpha}(x)
}+
\displaystyle{
\sum_{i=1}^{n+1}d_{i-1} \,
{\frac{i}{n+1}}\, \Bezier{i}{n+1}{\alpha}(x)
}\\
&=&
\displaystyle{
d_0\,\Bezier{0}{n+1}{\alpha}(x) +
\sum_{i=1}^{n}
\left[
{\frac{n+1-i}{n+1}}\, d_{i} +{\frac{i}{n+1}}\, d_{i-1} 
\right]
\Bezier{i}{n+1}{\alpha}(x)+
d_n\, \Bezier{n+1}{n+1}{\alpha}(x)
}\\
\end{array}
$$
This proves the result. 

\begin{castest}
For illustration we show here the  B\'ezier curves $\displaystyle{B_\alpha}$ 
of degree $3$ for  $\displaystyle{ \alpha \in \lbrace -1, 2, 5, \infty \rbrace}$. 
The  control polygons of the respective curves are as follow :
$$
\begin{array}{rcl}
\displaystyle{\Pi_{a}}
&=&
\displaystyle{
\lbrace
\couple{0}{2}, \couple{ 3.5}{0}, \couple{3.5}{4}, \couple{0}{0}
\rbrace
}
\\
\displaystyle{\Pi_{b}}
&=&
\displaystyle{
\lbrace
\couple{0}{1}, \couple{3.5}{0}, \couple{3.5}{4}, \couple{0}{1}
\rbrace
}
\\
\displaystyle{\Pi_{c}}
&=&
\displaystyle{
\lbrace
\couple{0}{1}, \couple{ 4.5}{4}, \couple{5.5}{0}, \couple{3.5}{1}
\rbrace
}
\\
\displaystyle{\Pi_{d}}
&=&
\displaystyle{
\lbrace
\couple{0}{1}, \couple{4}{0.5}, \couple{2.5}{3}, \couple{6}{3}
\rbrace
}
\\
\displaystyle{\Pi_{e}}
&=&
\displaystyle{
\lbrace
\couple{0}{1.5}, \couple{4}{0.5}, \couple{5}{4}, \couple{3}{2}
\rbrace
}
\\
\displaystyle{\Pi_{f}}
&=&
\displaystyle{
\lbrace
\couple{0}{3.5}, \couple{4}{0.5}, \couple{5}{4}, \couple{0}{0}
\rbrace
}
\\
\displaystyle{\Pi_{g}}
&=&
\displaystyle{
\lbrace
\couple{0}{3.5}, \couple{4}{0.5}, \couple{4.5}{2.5}, \couple{0}{0}
\rbrace
}
\\
\displaystyle{\Pi_{h}}
&=&
\displaystyle{
\lbrace
\couple{0}{0}, \couple{2}{2.5}, \couple{4.5}{3}, \couple{6.5}{1.5}
\rbrace
}
\\
\displaystyle{\Pi_{i}}
&=&
\displaystyle{
\lbrace
\couple{0}{3.5}, \couple{5}{1}, \couple{5}{1}, \couple{0}{0}
\rbrace
}
\\
\end{array}
$$

\begin{figure}[h!]
\begin{center}
\includegraphics[width=11cm]{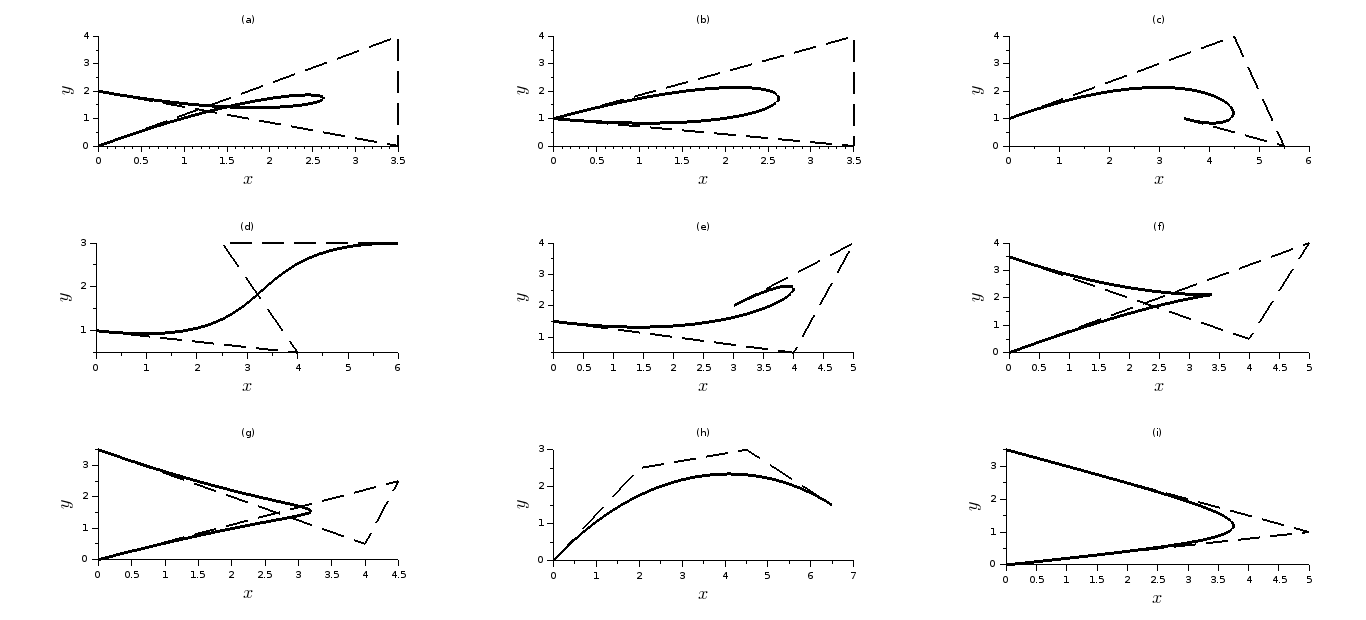}
\caption{The curves $\displaystyle{B_{-1}}$}
\label{figCourbeAlfa1}
\end{center}
\end{figure}

\begin{figure}[h!]
\begin{center}
\includegraphics[width=11cm]{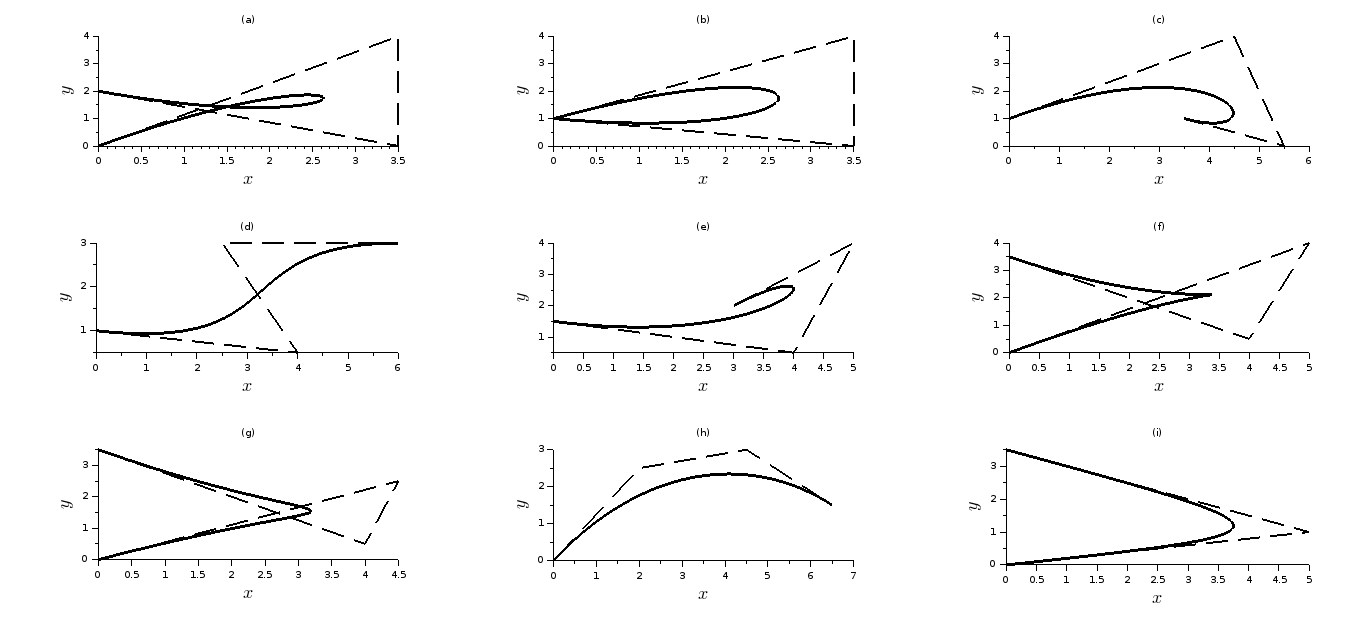}
\caption{The curves  $\displaystyle{B_{2}}$}
\label{figCourbeAlfa2}
\end{center}
\end{figure}

\begin{figure}[h!]
\begin{center}
\includegraphics[width=11cm]{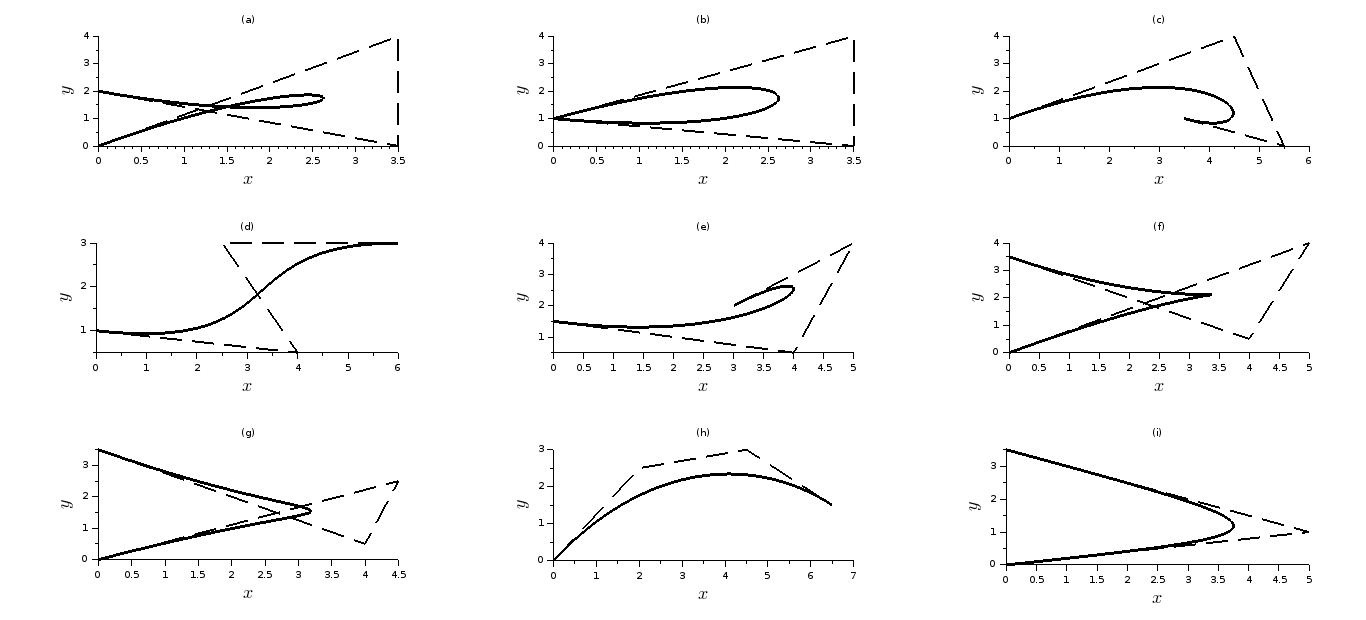}
\caption{The curves  $\displaystyle{B_{5}}$}
\label{figCourbeAlfa3}
\end{center}
\end{figure}

\begin{figure}[h!]
\begin{center}
\includegraphics[width=11cm]{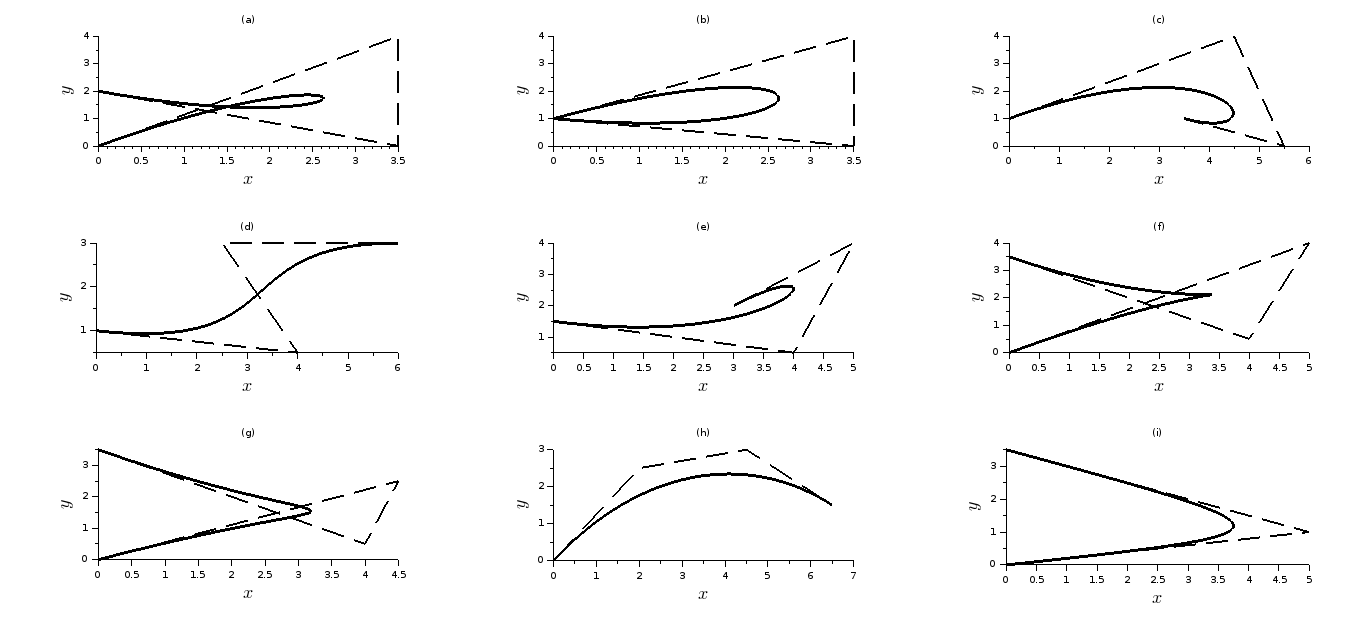}
\caption{The curves  $\displaystyle{B_{\infty}}$}
\label{figCourbeAlfa4}
\end{center}
\end{figure}

A comparative  analysis of  figures ~\ref{figCourbeAlfa1}, ~\ref{figCourbeAlfa2},
~\ref{figCourbeAlfa3},  and  \ref{figCourbeAlfa4} suggest that the   
B\'ezier curves are independent in index $\alpha$, but each curve is  function of its 
control polygon.
\end{castest}


\subsection{Algorithms for computing the B\'ezier curve}
These algorithms show that it is possible to calculate a point of the B\'ezier curve or all without the need to build the basis of Bernstein. 
The most fondemental is deCasteljau algorithm which can be formulate by
the following proposition :
\begin{proposition}[deCasteljau algorithm] \label{algoCasteljau}
For all  $r=0, \ldots , n$ and
  $x \in [a\, , \, b]$ we have : 
$$
\displaystyle{B_\alpha(x) = 
\sum_{i=0}^{n-r}d_{i}^{r}(x) \Bezier{i}{n-r}{\alpha}(x) }
$$
with
$$
\left\lbrace
\begin{array}{ll}
d_{i}^{0}(x)=d_i & \forall i=0, \ldots , n\\
\\
\displaystyle{
d_{i}^{r+1}(x)=w(x) d_{i+1}^{r}(x) +(1-w(x)) d_{i}^{r}(x)}
&\forall r=0, \ldots , n-1\\
&\forall i=0, \ldots , n-r-1\\
\end{array}
\right.
$$
where
$\displaystyle{w(x) = f_\alpha (x)}$

Moreover we have 
$\displaystyle{B_\alpha(x) = d_{0}^{n}(x)  }$
\end{proposition}

\Preuve
As in \cite{duncan2005}, we proceed by induction to prove this
algorithm.

Let $x\in [a\, ,\, b]$ and $w(x)= f_\alpha (x)$.

Since
$
\displaystyle{
\Bezier{i}{n}{\alpha}(x) =
w(x)\Bezier{i-1}{n-1}{\alpha}(x)  +
{\left(1- w(x) \right)}\Bezier{i}{n-1}{\alpha}(x) 
}
$
and
$
\displaystyle{
\Bezier{-1}{n}{\alpha}(x) =
\Bezier{n+1}{n}{\alpha}(x) \equiv 0
}
$
then
$$
\begin{array}{rcl}
\displaystyle{B_\alpha(x)} 
&= &
\displaystyle{
\sum_{i=0}^{n}d_{i}
\Bezier{i}{n}{\alpha}(x) 
}\\
&= &
\displaystyle{
\sum_{i=0}^{n}d_{i}
{\left[
w(x)\Bezier{i-1}{n-1}{\alpha}(x)  +
{\left(1- w(x) \right)}\Bezier{i}{n-1}{\alpha}(x) 
\right]}
}\\
&= &
\displaystyle{
\sum_{i=0}^{n}d_{i}
w(x)\Bezier{i-1}{n-1}{\alpha}(x)  +
\sum_{i=0}^{n}d_{i} 
{\left(1- w(x) \right)}\Bezier{i}{n-1}{\alpha}(x) 
}\\
&= &
\displaystyle{
\sum_{i=0}^{n-1}
{\left[
 w(x) d_{i+1} +
{\left(1- w(x) \right)} d_{i}
\right]}
\Bezier{i}{n-1}{\alpha}(x)
=
\sum_{i=0}^{n-1}
 d_{i}^{1}(x) \Bezier{i}{n-1}{\alpha}(x)
}\\
\end{array}
$$
with
$
\displaystyle{
 d_{i}^{1}(x)=
 {\left(1- w(x) \right)} d_{i}^{0}(x) +
 w(x) d_{i+1}^{0}(x)
}
$
where
$\displaystyle{
 d_{i}^{0}(x)= d_{i} \quad \forall i
}
$

Now we will show that  for all  $1\leq r \leq n$  we have
$
\displaystyle{B_\alpha(x)} 
=
\displaystyle{
\sum_{i=0}^{n-r}d_{i}^{r}
\Bezier{i}{n-r}{\alpha}(x) 
}
$
with 
$
\displaystyle{
 d_{i}^{r}(x)=
 {\left(1- w(x) \right)} d_{i}^{r-1}(x) +
 w(x) d_{i+1}^{r-1}(x)
}
$

Assume that for  $1\leq r < n$, we have
$
\displaystyle{B_\alpha(x)} 
=
\displaystyle{
\sum_{i=0}^{n-r}d_{i}^{r}
\Bezier{i}{n-r}{\alpha}(x) 
}
$  
with \\
$
\displaystyle{
 d_{i}^{r}(x)=
 {\left(1- w(x) \right)} d_{i}^{r-1}(x) +
 w(x) d_{i+1}^{r-1}(x)
}
$
Then
$$
\begin{array}{rcl}
\displaystyle{B_\alpha(x)} 
&= &
\displaystyle{
\sum_{i=0}^{n-r}d_{i}^{r}
\Bezier{i}{n-r}{\alpha}(x) 
}\\
&= &
\displaystyle{
\sum_{i=0}^{n-r}d_{i}^{r}
{\left[
w(x)\Bezier{i-1}{n-r-1}{\alpha}(x)  +
{\left(1- w(x) \right)}\Bezier{i}{n-r-1}{\alpha}(x) 
\right]}
}\\
&= &
\displaystyle{
\sum_{i=0}^{n-r}d_{i}^{r}
w(x)\Bezier{i-1}{n-r-1}{\alpha}(x)  +
\sum_{i=0}^{n-r}d_{i}^{r}
{\left(1- w(x) \right)}\Bezier{i}{n-r-1}{\alpha}(x) 
}\\
&= &
\displaystyle{
\sum_{i=0}^{n-r-1}d_{i+1}^{r}
w(x)\Bezier{i}{n-r-1}{\alpha}(x)  +
\sum_{i=0}^{n-r-1}d_{i}^{r}
{\left(1- w(x) \right)}\Bezier{i}{n-r-1}{\alpha}(x) 
}\\
&= &
\displaystyle{
\sum_{i=0}^{n-r-1}
{\left[
w(x)d_{i+1}^{r}+
{\left(1- w(x) \right)}d_{i}^{r}
\right]}
\Bezier{i}{n-r-1}{\alpha}(x) 
=
\sum_{i=0}^{n-r-1}d_{i}^{r+1}
\Bezier{i}{n-r-1}{\alpha}(x) 
}\\
\end{array}
$$
with 
$
\displaystyle{
 d_{i}^{r+1}(x)=
 {\left(1- w(x) \right)} d_{i}^{r}(x) +
 w(x) d_{i+1}^{r}(x)
}
$
and we deduce the result.

\begin{proposition}[subdivision algorithm]\label{algoSubdiv}
Let $c \in ]a\, , \, b[$. The  point  
$\displaystyle{B_\alpha(c) = d_{0}^{n}(c)  }$  computed by the  relation
$$
\left\lbrace
\begin{array}{ll}
d_{i}^{0}(c)=d_i & \forall i=0, \ldots , n\\
\\
\displaystyle{
d_{i}^{r+1}(c)=w(c) d_{i+1}^{r}(c) +(1-w(c)) d_{i}^{r}(c)}
&\forall r=0, \ldots , n-1\\
&\forall i=0, \ldots , n-r-1\\
\end{array}
\right.
$$
where 
$\displaystyle{w(x) = f_\alpha (x)}$
divides the curve $B_\alpha$ in two  rational B\'ezier curves
of  $\alpha$ index and degree $n$ on the  parameter space 
$[a\, , \, b]$  whose the control polygon points are   respectively
$\displaystyle{\suite{d_{0}^{i}(c)}_{i=0}^{n}}$
and
$\displaystyle{\suite{d_{i}^{n-i}(c)}_{i=0}^{n}}$

\end{proposition}

The proof of this  proposition uses the following three lemmas  :

\begin{lemme}\label{algoSubdivLem1}
The points $\displaystyle{\suite{d_{k}^{j}(c)}}$ defined by the algorithm of 
the  proposition~ \ref{algoSubdiv}, satisty for all  $0\leq k \leq j\leq n$, 
the  relation
$$
\displaystyle{
d_{k}^{j}(c) = 
\sum_{i=0}^{j} d_{i+k} \Bezier{i}{j}{\alpha}(c)
}
$$
\end{lemme}

\Preuve
Let $c \in ]a\, , \, b[$ and  $\displaystyle{ w=f_\alpha (c)}$.
From the  proposition~ \ref{algoSubdiv}, for all  $0\leq k \leq j\leq n$, 
we have 
$
\displaystyle{
d_{k}^{j}(c) =
{\left(1- w \right)} d_{k}^{j-1}(c) + w d_{k+1}^{j-1}(c) 
}
$.
we proceed by induction to prove that  for all  $1 \leq r \leq j$, 
$
\displaystyle{
d_{k}^{j}(c) =
\sum_{i=0}^{r} d_{k+i}^{j-r}(c) \Bezier{i}{r}{\alpha}(c)
}
$ 
and we deduce the result  for  $r=j$.

For all  $j \geq 1$ we have
$$
\begin{array}{rcl}
\displaystyle{
d_{k}^{j}(c)
}
&=&
\displaystyle{
{\left(1- w \right)} d_{k}^{j-1}(c) + w d_{k+1}^{j-1}(c) 
}
\\
&=&
\displaystyle{
\Bezier{0}{1}{\alpha}(c) d_{k}^{j-1}(c) + 
\Bezier{1}{1}{\alpha}(c) d_{k+1}^{j-1}(c) 
}
\\
&=&
\displaystyle{
\sum_{i=0}^{1}\Bezier{i}{r=1}{\alpha}(c) d_{k+i}^{j-1}(c) 
}
\\
\end{array}
$$

Assume that for all  $1\leq r < j$, we have
$
\displaystyle{
d_{k}^{j}(c) =
\sum_{i=0}^{r} d_{k+i}^{j-r}(c) \Bezier{i}{r}{\alpha}(c)
}
$
then using the  proposition~\ref{DegrePlus}, we have 
$$
\begin{array}{rcl}
\displaystyle{
d_{k}^{j}(c)
}
&=&
\displaystyle{
\sum_{i=0}^{r} d_{k+i}^{j-r}(c) \Bezier{i}{r}{\alpha}(c)
}
\\
&=&
\displaystyle{
\sum_{i=0}^{r} 
\left[
(1-w) d_{k+i}^{j-r-1}(c) +
w d_{k+i+1}^{j-r-1}(c) 
\right]
\Bezier{i}{r}{\alpha}(c)
}
\\
&=&
\displaystyle{
\sum_{i=0}^{r} 
 d_{k+i}^{j-r-1}(c)
\left[
(1-w)
\Bezier{i}{r}{\alpha}(c)
\right]

}
+
\displaystyle{
\sum_{i=0}^{r} 
d_{k+i+1}^{j-r-1}(c) 
\left[
w 
\Bezier{i}{r}{\alpha}(c)
\right]
}
\\
&=&
\displaystyle{
\sum_{i=0}^{r} 
 d_{k+i}^{j-r-1}(c)
\left[
\frac{r+1-i}{r+1}
\Bezier{i}{r+1}{\alpha}(c)
\right]

}
+
\displaystyle{
\sum_{i=0}^{r} 
d_{k+i+1}^{j-r-1}(c) 
\left[
\frac{i+1}{r+1}
\Bezier{i+1}{r+1}{\alpha}(c)
\right]
}
\\
&=&
\displaystyle{
\sum_{i=0}^{r} 
 d_{k+i}^{j-r-1}(c)
\left[
\frac{r+1-i}{r+1}
\Bezier{i}{r+1}{\alpha}(c)
\right]

}
+
\displaystyle{
\sum_{i=1}^{r+1} 
d_{k+i}^{j-r-1}(c) 
\left[
\frac{i}{r+1}
\Bezier{i}{r+1}{\alpha}(c)
\right]
}
\\
&=&
\displaystyle{
\sum_{i=0}^{r+1} 
 d_{k+i}^{j-r-1}(c)
\Bezier{i}{r+1}{\alpha}(c)
}
\\
\end{array}
$$
This completes the proof.

\begin{lemme}\label{algoSubdivLem2}
Let  $\displaystyle{\alpha \in ]-\infty \, , \, 0[ \cup ]1\, , \, \infty[}$,
and $a,\, b \in \RR$ with $a<b$. Let $c\in ]a\, , \, b[$ and 
$\displaystyle{f_\alpha \in {\cal H}([a\, , \, b])}$.
There exists  a bijective mapping 
$u$ from $[a\, , \, b]$ to  $[a\, , \, c]$ such that 
$\displaystyle{f_\alpha \circ u =f_\alpha (c) f_\alpha }$

Moreover,  for all  $n \in \NN^*$ and $i \in \NN$ with $i\leq n$ we have 
$$
\displaystyle{
\Bezier{i}{n}{\alpha}(u(t))=
\sum_{j=0}^{n}\Bezier{i}{j}{\alpha}(c)\Bezier{j}{n}{\alpha}(t)
},
\,
\forall t \in [a\, , \, b]
$$
\end{lemme}

\Preuve
\noindent
\emph{Step 1}
First of all let us prove the existence of $u$

For  $c\in ]a\, , \, b[$, we have  $f_\alpha (c) \in ]0\, , \, 1[$. For all   $t \in [a\, , \, b]$, we have
$f_\alpha (t) \in [0\, , \, 1]$. Since $f_\alpha$ is increasing strictly then 
$$
\displaystyle{
w=f_\alpha (c) f_\alpha (t) \in [0\, , \, f_\alpha (c) ]
=f_\alpha ([a\, , \, c]) \subset [0 \, , \, 1]
}
$$
Using the fact that   $f_\alpha$ is  a bijection we have an unique
$u(t) \in [a\, , \, c]$ such that
$
\displaystyle{
f_\alpha (u(t)) = w = f_\alpha (c) f_\alpha (t)
}
$.

This implies the  mapping
$
\displaystyle{
t\in [a\, , \, b]
\mapsto
u(t) =f_\alpha^{-1}
\left(
 f_\alpha (c) f_\alpha (t)
\right)
\in  [a\, , \, c]
}
$

Let  $x, y \in  [a\, , \, c]$ such that $x < y$, we have
$
\displaystyle{
f_\alpha (x) <  f_\alpha (y)
}
$.
Then 
$
\displaystyle{
f_\alpha (c)f_\alpha (x) < f_\alpha (c) f_\alpha (y)
}
$
and
$$
u(x) =f_\alpha^{-1}
\left(
 f_\alpha (c) f_\alpha (x)
\right)
<
f_\alpha^{-1}
\left(
 f_\alpha (c) f_\alpha (y)
\right)
=u(y)
$$
$u$ is increasing strictly. Then $u$ is  an increasing  injection.
$$
\begin{array}{rcl}
u(a) &=&
\displaystyle{
f_\alpha^{-1}
\left(
 f_\alpha (c) f_\alpha (a)
\right)
=
f_\alpha^{-1}(0) =a
}\\
u(b) &=&
\displaystyle{
f_\alpha^{-1}
\left(
 f_\alpha (c) f_\alpha (b)
\right)
=
f_\alpha^{-1}
\left(
 f_\alpha (c) 
\right) = c
}\\
\end{array}
$$

Since $f_\alpha$ and $f_\alpha^{-1}$ are continuous then $u$ is  continuous and we have
$$
\displaystyle{
u{\left([a\, , \, b]\right)} = [u(a)\, , \, u(b)]= [a\, , \, c]
}
$$

Therefore $u$ is a bijection from  $[a\, , \, b]$ to  $[a\, , \, c]$ and satisfies
$\displaystyle{f_\alpha \circ u =f_\alpha (c) f_\alpha }$.

\noindent
\emph{Step 2}

Let $t \in [a\, , \, b]$, $\displaystyle{w(t)=f_\alpha (t)}$ and
 $\displaystyle{w(c)=f_\alpha (c)}$ we have
 $$
\begin{array}{l}
 \displaystyle{\bar{w}=f_\alpha \circ u (t) =w(c)w(t)}\\
\displaystyle{1 -\bar{w}=1-w(c)w(t)=\left(1-w(t)\right)+w(t)\left(1-w(c)\right)}\\
\end{array}
 $$
We deduce
$$
\begin{array}{rcl}
{\left(
1-\bar{w}
\right)}^{n-i}
&=&
\displaystyle{
{\left[
\left(1-w(t)\right)+w(t)\left(1-w(c)\right)
\right]}^{n-i}
}
\\
&=&
\displaystyle{
\sum_{j=0}^{n-i}
C_{n-i}^{j}
{\left(1-w(t)\right)}^{j}
{\left[
w(t)\left(1-w(c)\right)
\right]}^{n-i-j}
}
\\
&&\\
\bar{w}^i
{\left(
1-\bar{w}
\right)}^{n-i}
&=&
\displaystyle{
\sum_{j=0}^{n-i}
C_{n-i}^{j}
{\left[w(t)w(c) \right]}^{i}
{\left(1-w(t)\right)}^{j}
{\left[
w(t)\left(1-w(c)\right)
\right]}^{n-i-j}
}
\\
&=&
\displaystyle{
\sum_{j=0}^{n-i}
C_{n-i}^{j}
{\left[w^{n-j}(t)
{\left(1-w(t)\right)}^{j}
 \right]}
{\left[
w^{i}(c)\left(1-w(c)\right)^{n-i-j}
\right]}
}
\\
\end{array}
$$

$$
\begin{array}{rcl}
\Bezier{i}{n}{\alpha}{\left(u(t)\right)}
&=&
C_{n}^{i}
\bar{w}^i
{\left(
1-\bar{w}
\right)}^{n-i}
\\
&=&
\displaystyle{
\sum_{j=0}^{n-i}
C_{n}^{i}
C_{n-i}^{j}
{\left[w^{n-j}(t)
{\left(1-w(t)\right)}^{j}
 \right]}
{\left[
w^{i}(c)\left(1-w(c)\right)^{n-i-j}
\right]}
}
\\
&=&
\displaystyle{
\sum_{j=0}^{n-i}
C_{n}^{i}
C_{n-i}^{j}
\frac{\Bezier{n-j}{n}{\alpha}(t)}{C_{n}^{n-j}}
\frac{\Bezier{i}{n}{\alpha}(c)}{C_{n}^{i}}
\left(1-w(c)\right)^{-j}
}
\\
&=&
\displaystyle{
\sum_{j=0}^{n-i}
\frac{C_{n-i}^{j}}{C_{n}^{n-j}}
\Bezier{n-j}{n}{\alpha}(t)
\Bezier{i}{n}{\alpha}(c)
\left(1-w(c)\right)^{-j}
}
\\
\end{array}
$$

By an iterative use of the  proposition~\ref{DegrePlus} we have
$$
\displaystyle{
\Bezier{i}{n}{\alpha}(c)
\left(1-w(c)\right)^{-j}
=\prod_{l=0}^{j-1}\left(\frac{n-l}{n-i-l}\right)
\Bezier{i}{n-j}{\alpha}(c)
=\frac{C_{n}^{j}}{C_{n-i}^{j}}
\Bezier{i}{n-j}{\alpha}(c)
}
$$

We deduce that
$$
\begin{array}{rcl}
\Bezier{i}{n}{\alpha}{\left(u(t)\right)}
&=&
\displaystyle{
\sum_{j=0}^{n-i}
\Bezier{n-j}{n}{\alpha}(t)
\Bezier{i}{n-j}{\alpha}(c)
}
\\
&=&
\displaystyle{
\sum_{j=i}^{n}
\Bezier{j}{n}{\alpha}(t)
\Bezier{i}{j}{\alpha}(c)
}
\\
&=&
\displaystyle{
\sum_{j=0}^{n}
\Bezier{j}{n}{\alpha}(t)
\Bezier{i}{j}{\alpha}(c)
}
\\
\end{array}
$$
 Since for all  $i>j$ we have
$
\displaystyle{
\Bezier{i}{j}{\alpha}(c)=0 
}
$,
the expected result follows.

\begin{lemme}\label{algoSubdivLem3}
Let $\displaystyle{\alpha \in ]-\infty \, , \, 0[ \cup ]1\, , \, \infty[}$,
and $a,\, b \in \RR$ and $a<b$. Let $c\in ]a\, , \, b[$ and  
$\displaystyle{f_\alpha \in {\cal H}([a\, , \, b])}$. There exists a bijective mapping 
$v$ from  $[a\, , \, b]$ to  $[c\, , \, b]$ such that for all   $t \in [a\, , \, b]$ 
$$
\displaystyle{f_\alpha \circ v(t) =1-{\left(1 - f_\alpha (c)\right)}
{\left( 1- f_\alpha (t)\right)}
}
$$

Moreover,  for all $n \in \NN^*$ and $i \in \NN$ with $i\leq n$
$$
\displaystyle{
\Bezier{i}{n}{\alpha}(v(t))=
\sum_{j=0}^{n}\Bezier{i-j}{n-j}{\alpha}(c)\Bezier{j}{n}{\alpha}(t)
},
\,
\forall t \in [a\, , \, b]
$$
\end{lemme}

\Preuve
\noindent
\emph{Step 1}
First of all we prove the  existence of $v$

Let $c\in ]a\, , \, b[$. We have  $f_\alpha (c) \in ]0\, , \, 1[$ since
$1- f_\alpha (c) \in ]0\, , \, 1[$. 
By same way, for all   $t \in [a\, , \, b]$, we have
$f_\alpha (t) \in [0\, , \, 1]$ and $1 - f_\alpha (t) \in [0\, , \, 1]$. 

Since $f_\alpha$ is increasing strictly
we deduce that,  for all   $t \in [a\, , \, b]$, we have an unique 
$$
\displaystyle{
w(t) =1-{\left(1- f_\alpha (c) \right)}{\left(1- f_\alpha (t) \right)}\in
[f_\alpha (c) \, , \, 1]
=f_\alpha ([c\, , \, b]) \subset [0\, , \, 1]
}
$$

Using the fact that $f_\alpha$ is a  bijection we have an unique
$v(t) \in [a\, , \, c]$ such that 
$
\displaystyle{
f_\alpha (v(t)) = w (t)
}
$.

This implies a mapping
$$
\displaystyle{
t\in [a\, , \, b]
\mapsto
v(t) =f_\alpha^{-1}
\left[
 1-{\left(1- f_\alpha (c) \right)}{\left(1- f_\alpha (t) \right)}
\right]
\in  [c\, , \, b]
}
$$ 

Let  $x, y \in  [c\, , \, b]$ such that $x < y$, we have
$
\displaystyle{
{\left(1- f_\alpha (c) \right)}{\left(1- f_\alpha (x) \right)}
>
{\left(1- f_\alpha (c) \right)}{\left(1- f_\alpha (y) \right)}
}
$.

Then
$$
\begin{array}{rcl}
v(x) 
&=&
\displaystyle{
f_\alpha^{-1}
\left[
 1-{\left(1- f_\alpha (c) \right)}{\left(1- f_\alpha (x) \right)}
\right]
}\\
&<&
\displaystyle{
f_\alpha^{-1}
\left[
 1-{\left(1- f_\alpha (c) \right)}{\left(1- f_\alpha (y) \right)}
\right]
}\\
&=&
v(y)\\
\end{array}
$$
$v$ is increasing strictly. $v$ is an increasing  injection from
$[a\, , \, b]$ to $[v(a)\, , \, v(b)]=[c\, , \, b]$ since $v$ is continuous.

Then $v$ is a bijection from $[a\, , \, b]$ to  $[c\, , \, b]$ and satisfies, 
for all  $t \in [a\, , \, b]$ 
$$
\displaystyle{
f_\alpha \circ v(t) =
 1-{\left(1- f_\alpha (c) \right)}{\left(1- f_\alpha (t) \right)}
}
$$

\noindent
\emph{Step 2}

Let $t \in [a\, , \, b]$, $\displaystyle{w(t)=f_\alpha (t)}$ and
 $\displaystyle{w(c)=f_\alpha (c)}$ we have
 $$
\begin{array}{rcl}
\displaystyle{\bar{w}}
&=&
\displaystyle{f_\alpha \circ v (t) }
\\
&=&
\displaystyle{1- \left(1-w(c)\right)\left(1-w(t)\right)}
\\
&=&
\displaystyle{w(t)+w(c)\left(1-w(t)\right)}
\\
\displaystyle{1 -\bar{w}}
&=&
\displaystyle{\left(1-w(t)\right)\left(1-w(c)\right)}\\
\end{array}
 $$

Then we have
$$
\begin{array}{rcl}
{\left(
\bar{w}
\right)}^{i}
&=&
\displaystyle{
{\left[
w(t)+w(c)\left(1-w(t)\right)
\right]}^{i}
}
\\
&=&
\displaystyle{
\sum_{j=0}^{i}
C_{i}^{j}
{\left(w(t)\right)}^{j}
{\left[
w(c)\left(1-w(t)\right)
\right]}^{i-j}
}
\\
&&\\
\bar{w}^i
{\left(
1-\bar{w}
\right)}^{n-i}
&=&
\displaystyle{
\sum_{j=0}^{i}
C_{i}^{j}
{\left[
{\left(w(t)\right)}^{j}
{\left(1-w(t)\right)}^{n-j}
\right]}
{\left[
{\left(w(c)\right)}^{i-j}
{\left(1-w(c)\right)}^{n-i}
\right]}
}
\\
\end{array}
$$

$$
\begin{array}{rcl}
\Bezier{i}{n}{\alpha}{\left(u(t)\right)}
&=&
C_{n}^{i}
\bar{w}^i
{\left(
1-\bar{w}
\right)}^{n-i}
\\
&=&
\displaystyle{
\sum_{j=0}^{i}
C_{n}^{i}
C_{i}^{j}
{\left[
{\left(w(t)\right)}^{j}
{\left(1-w(t)\right)}^{n-j}
\right]}
{\left[
{\left(w(c)\right)}^{i-j}
{\left(1-w(c)\right)}^{n-i}
\right]}
}
\\
&=&
\displaystyle{
\sum_{j=0}^{i}
C_{i}^{j}
{\left[
{\left(w(t)\right)}^{j}
{\left(1-w(t)\right)}^{n-j}
\right]}
{\left[
C_{n}^{i}
{\left(w(c)\right)}^{i-j}
{\left(1-w(c)\right)}^{n-i}
\right]}
}
\\
&=&
\displaystyle{
\sum_{j=0}^{i}
C_{i}^{j}
\frac{\Bezier{j}{n}{\alpha}(t)}{C_{n}^{j}}
\Bezier{i}{n}{\alpha}(c)
\left(w(c)\right)^{-j}
}
\\
&=&
\displaystyle{
\sum_{j=0}^{i}
\frac{C_{i}^{j}}{C_{n}^{j}}
\Bezier{j}{n}{\alpha}(t)
\Bezier{i}{n}{\alpha}(c)
\left(w(c)\right)^{-j}
}
\\
\end{array}
$$

An iterative use of the  proposition~\ref{DegrePlus} implies
$$
\displaystyle{
\Bezier{i}{n}{\alpha}(c)
\left(w(c)\right)^{-j}
=\prod_{l=0}^{j-1}\left(\frac{n-l}{i-l}\right)
\Bezier{i}{n-j}{\alpha}(c)
=\frac{C_{n}^{j}}{C_{i}^{j}}
\Bezier{i}{n-j}{\alpha}(c)
}
$$

We deduce that 
$$
\begin{array}{rcl}
\Bezier{i}{n}{\alpha}{\left(v(t)\right)}
&=&
\displaystyle{
\sum_{j=0}^{i}
\Bezier{j}{n}{\alpha}(t)
\Bezier{i-j}{n-j}{\alpha}(c)
}
\\
&=&
\displaystyle{
\sum_{j=i}^{n}
\Bezier{i-j}{n-j}{\alpha}(c)
\Bezier{j}{n}{\alpha}(t)
}
\\
\end{array}
$$
because that, for all  $i<j$ we have 
$
\displaystyle{
\Bezier{n-j}{i-j}{\alpha}(c)=0 
}
$

\Preuve 
\emph{Proof of the  proposition~\ref{algoSubdiv}}

Let  $\bar{B}_\alpha$ and  $\hat{B}_\alpha$ be the  restrictions 
of  $B_\alpha$ on the  respective intervals \\
$ [a\, , \, c] \subset  [a\, , \, b]$ and 
$ [c\, , \, b] \subset  [a\, , \, b]$ 
 mapping the parameter space $ [a\, , \, b]$.

For all  $t \in  [a\, , \, b]$, using the  lemma~\ref{algoSubdivLem2} we have
$u(t)  \in  [a\, , \, c]$ and
$$
\begin{array}{rcl}
\displaystyle{
\bar{B}_\alpha (t)
}
&=&
\displaystyle{
B_\alpha (u(t))
}
\\
&=&
\displaystyle{
\sum_{i=0}^{n} d_{i}\Bezier{i}{n}{\alpha}(u(t))
}
\\
&=&
\displaystyle{
\sum_{i=0}^{n} d_{i}
\left[
\sum_{j=0}^{n}\Bezier{i}{j}{\alpha}(c)\Bezier{j}{n}{\alpha}(t)
\right]
}
\\
&=&
\displaystyle{
\sum_{i=0}^{n}
\left[
\sum_{j=0}^{n}
\left(
 d_{i} \Bezier{i}{j}{\alpha}(c)
\right)
\Bezier{j}{n}{\alpha}(t)
\right]
}
\\
&=&
\displaystyle{
\sum_{j=0}^{n}
\left[
\sum_{i=0}^{n}
 d_{i} \Bezier{i}{j}{\alpha}(c)
\right]
\Bezier{j}{n}{\alpha}(t)
}
\\
&=&
\displaystyle{
\sum_{j=0}^{n}
 d_{0}^{j}(c) \Bezier{j}{n}{\alpha}(t)
}
\\
\end{array}
$$
That is the expected  result by using the  lemma~\ref{algoSubdivLem1}

With the  lemma~\ref{algoSubdivLem3}, for all  $t \in [a\, ,\, b]$  we have
$$
\begin{array}{rcl}
\hat{B}_\alpha (t)
&=&
B_\alpha {\left(v(t)\right)}
\\
&=&
\displaystyle{
\sum_{i=0}^{n} d_{i}
\Bezier{i}{n}{\alpha} {\left(v(t)\right)}
}
\\
&=&
\displaystyle{
\sum_{i=0}^{n} d_{i}
{\left(
\sum_{j=0}^{n} 
\Bezier{i-j}{n-j}{\alpha}(c) 
\Bezier{j}{n}{\alpha}(t)
\right)}
}
\\
&=&
\displaystyle{
\sum_{i=0}^{n}
{\left(
\sum_{j=0}^{n} d_{i} 
\Bezier{i-j}{n-j}{\alpha}(c) 
\Bezier{j}{n}{\alpha}(t)
\right)}
}
\\
&=&
\displaystyle{
\sum_{j=0}^{n}
{\left(
\sum_{i=0}^{n} d_{i} 
\Bezier{i-j}{n-j}{\alpha}(c) 
\Bezier{j}{n}{\alpha}(t)
\right)}
}
\\
&=&
\displaystyle{
\sum_{j=0}^{n}
{\left(
\sum_{i=0}^{n} d_{i} 
\Bezier{i-j}{n-j}{\alpha}(c) 
\right)}
\Bezier{j}{n}{\alpha}(t)
}
\\
&=&
\displaystyle{
\sum_{j=0}^{n}
{\left(
\sum_{i=j}^{n} d_{i} 
\Bezier{i-j}{n-j}{\alpha}(c) 
\right)}
\Bezier{j}{n}{\alpha}(t)
}
\\
&=&
\displaystyle{
\sum_{j=0}^{n}
{\left(
\sum_{i=0}^{n-j} d_{i+j} 
\Bezier{i}{n-j}{\alpha}(c) 
\right)}
\Bezier{j}{n}{\alpha}(t)
}
\\
&=&
\displaystyle{
\sum_{j=0}^{n} d_{j}^{n-j}
\Bezier{j}{n}{\alpha}(t)
}
\\
\end{array}
$$
using the  lemma~\ref{algoSubdivLem1}.

\begin{corollaire}[Algorithm of B\'ezier curve construction ]
Let $c =\displaystyle{\frac{a+b}{2}}$. The  point  
$\displaystyle{B_\alpha(c) = d_{0}^{n}  }$ computed by the  relation
$$
\left\lbrace
\begin{array}{ll}
d_{i}^{0}=d_i & \forall i=0, \ldots , n\\
\\
\displaystyle{
d_{i}^{r+1}=\frac{\alpha}{2\alpha-1} d_{i+1}^{r} +
\frac{\alpha-1}{2\alpha-1} d_{i}^{r}}
&\forall r=0, \ldots , n-1\\
&\forall i=0, \ldots , n-r-1\\
\end{array}
\right.
$$
subdivides the rational B\'ezier curve $B_\alpha$ in two rational B\'ezier curves 
of $\alpha$ index and degree $n$, on the  parameter space $[a\, , \, b]$  
with the respective control polygon points :  
$\displaystyle{\suite{d_{0}^{i}}_{i=0}^{n}}$
and
$\displaystyle{\suite{d_{i}^{n-i}}_{i=0}^{n}}$

\end{corollaire}

\Preuve
A direct application of the  proposition~\ref{algoSubdiv}.


\begin{castest}

The  figures \ref{figSubdivision1},  \ref{figSubdivision2},  \ref{figSubdivision3}
and  \ref{figSubdivision4}
point out some alternatives of the  subdivision algorithm iterate on  4 levels
in the  construction of the B\'ezier curve $B_\alpha$ with  control  polygon 
$
\displaystyle{\Pi}
=
\displaystyle{
\lbrace
\couple{0}{3.5}, \couple{4}{0.5}, \couple{4.5}{2.5}, \couple{0}{0}
\rbrace
}
$ 
with $\alpha \in \lbrace -1, 2, 5, \infty \rbrace$

\begin{figure}[h!]
\begin{center}
\includegraphics[width=11cm]{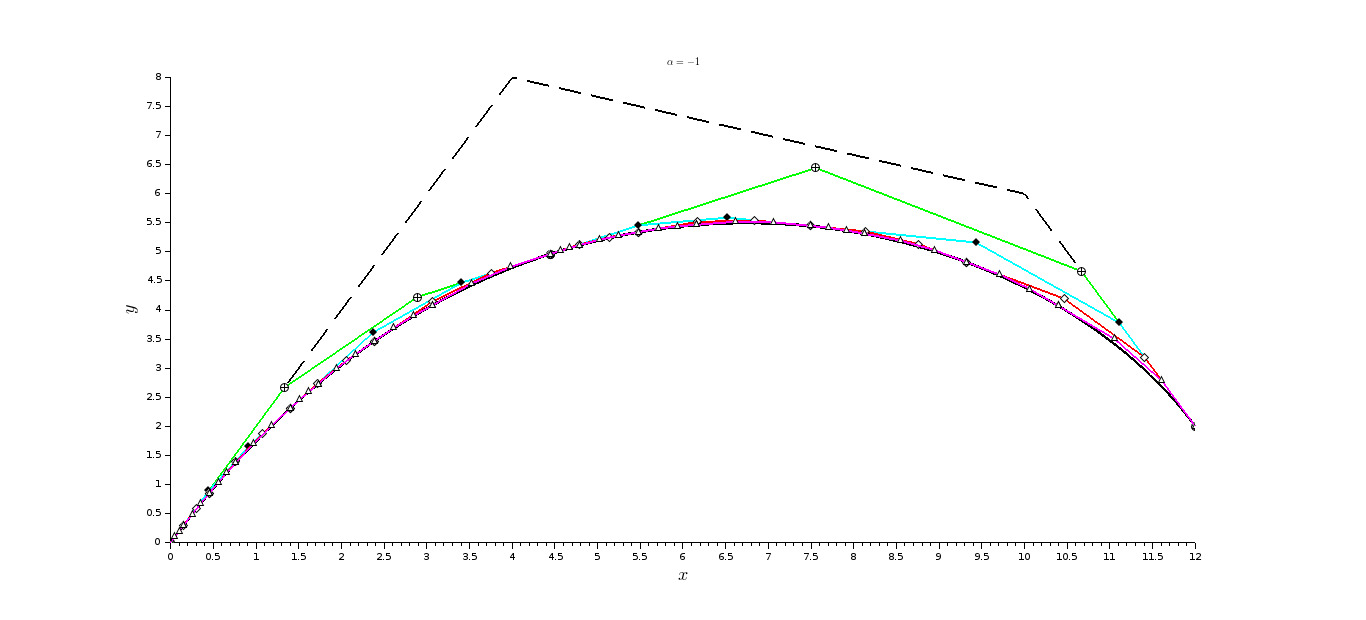}
\caption{B\'ezier curve $\displaystyle{B_{-1}}$ and control 
sub-polygons on levels up to $4$}
\label{figSubdivision1}
\end{center}
\end{figure}

\begin{figure}[h!]
\begin{center}
\includegraphics[width=11cm]{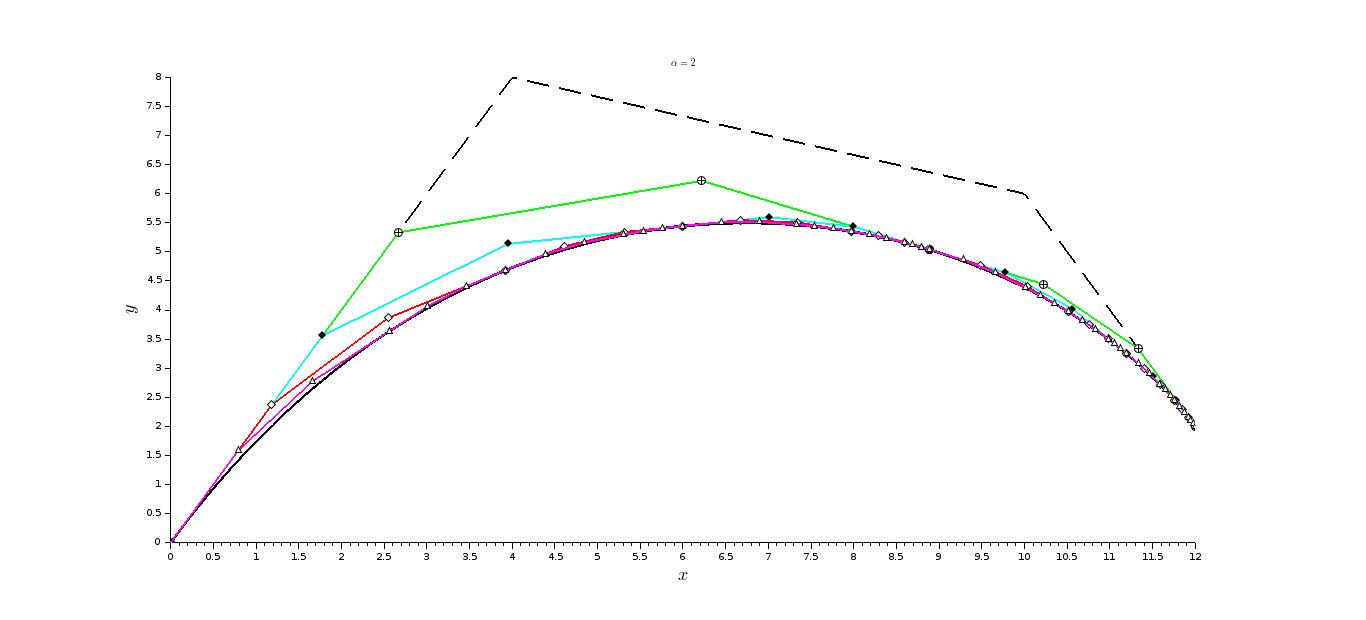}
\caption{B\'ezier curve $\displaystyle{B_{2}}$ and control 
sub-polygons on levels up to $4$}
\label{figSubdivision2}
\end{center}
\end{figure}

\begin{figure}[h!]
\begin{center}
\includegraphics[width=11cm]{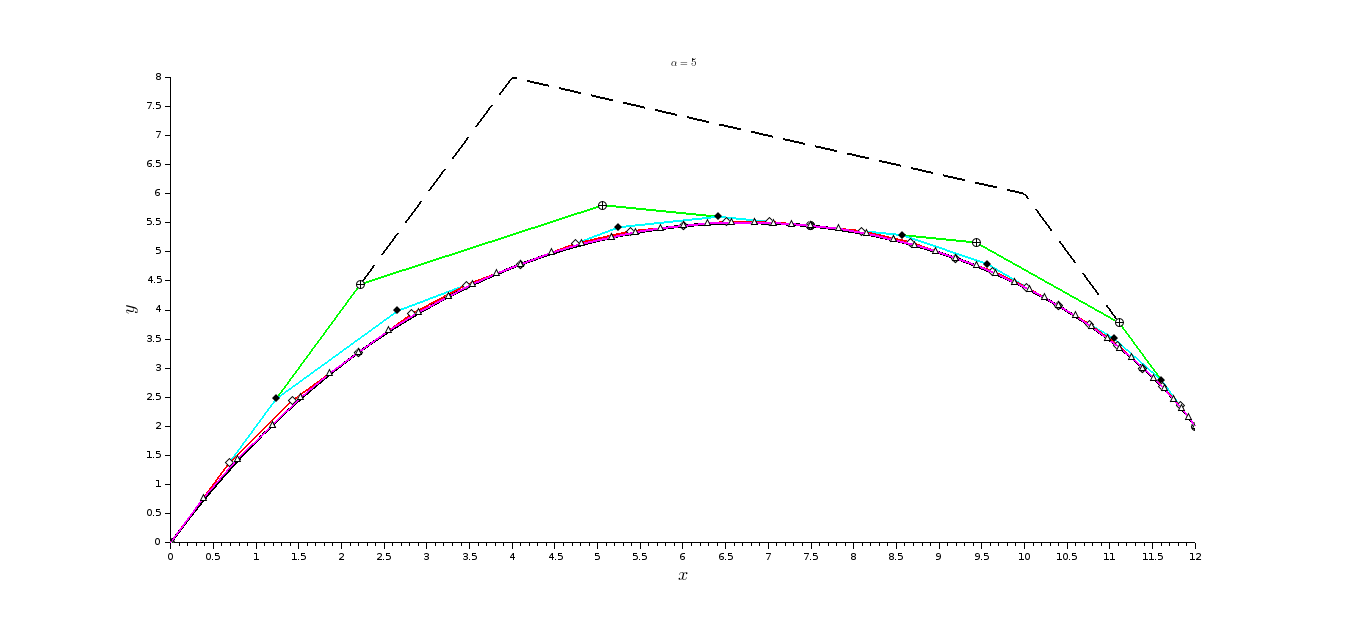}
\caption{B\'ezier curve $\displaystyle{B_{5}}$ and control 
sub-polygons on levels up to $4$}
\label{figSubdivision3}
\end{center}
\end{figure}

\begin{figure}[h!]
\begin{center}
\includegraphics[width=11cm]{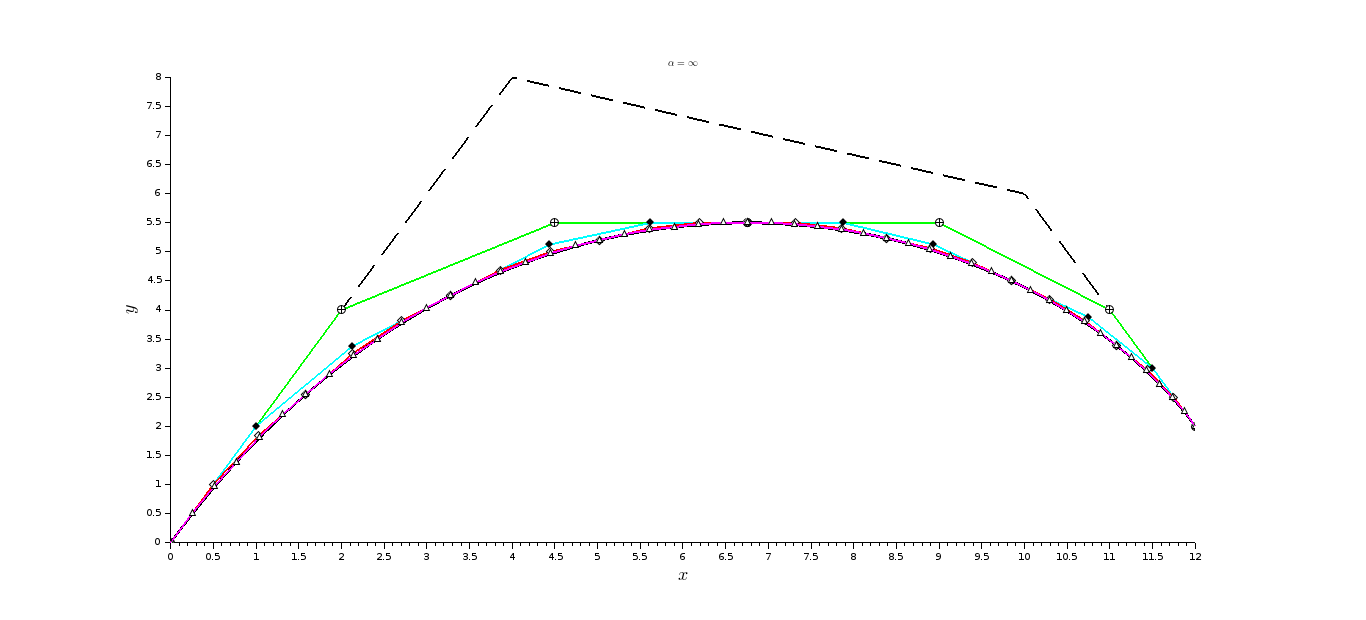}
\caption{B\'ezier curve $\displaystyle{B_{\infty}}$ and control 
sub-polygons on levels up to $4$}
\label{figSubdivision4} 
\end{center}
\end{figure}

From the analysis of the figures  \ref{figSubdivision1} and  \ref{figSubdivision2} 
we can observe that the chronological repartition  of the
control sub-polygons points of the B\'ezier curves $B_{-1}$ et  $B_{2}$, 
points out the symmetrical effect of the indexes $\alpha$ and  $1-\alpha$. 
In the B\'ezier curve $B_{-1}$ the  points concentration converge to the endpoint of
parameter $0$, 
in the B\'ezier curve  $B_{2}$  the  points concentration converge to 
the endpoint of parameter  $1$.

The  points  equirepartition in the figure \ref{figSubdivision4} confirms
the  self symmetry shown in the  standard  case which corresponds
to $\alpha=\infty$.
\end{castest}

\begin{remarque}
In conducting evidence geometric properties B\'ezier curves discussed above, 
 the nature of the functions homographic
$\displaystyle{f_\alpha \in {\cal H}{([a\, , \, b])}}$  was not exploited. 
We only used the fact that are an increasing bijection of 
$[a\, , \, b]$ into  $[0\, , \, 1]$.

The following proposition shows that the curves  $B_\alpha$ are completely defined 
by their points of control. Which reinforces this observation.
In this proof we only used the fact that $f_\alpha$
is a  diffeomorphism of the  class ${\cal C}^2$.
\end{remarque}

\begin{proposition}
Let $n, \, d \in \NN^*$ and we 
consider $n+1$ points $\displaystyle{\suite{d_i}_{i=0}^{n}\in \RR^d}$.
Let $\displaystyle{\alpha, \, \beta \in ]-\infty \, , \, 0[\cup ]1 \, , \, \infty [}$ and
$\displaystyle{a,\, b,\,a_1,\, b_1\,\in \RR}$ such that $a<b$ and  $a_1<b_1$.
The rational B\'ezier curve $B_\alpha$ of $\alpha$ index and degree $n$ 
with  control polygon points $\displaystyle{\suite{d_i}_{i=0}^{n}}$ on
 the parameter space $[a\, , \, b]$ is same to 
 the rational B\'ezier curve  $B_\beta$ of $\beta$  index and degree $n$ 
with  control polygon points 
$\displaystyle{\suite{d_i}_{i=0}^{n}}$,
  the parameter space $[a_1\, , \, b_1]$
 
\end{proposition}

\Preuve
Let $M\in B_\alpha$ and $f_\alpha \in {\cal H}{([a\, , \, b])}$, 
then there exists $x \in [a\, , \, b]$ such that
$\displaystyle{w=f_\alpha (x)\in [0\, , \, 1] }$ and 
$\displaystyle{M=B_\alpha (x) = \sum_{i=0}^{n} d_{i}\Bezier{i}{n}{\alpha} (x)}$
with
$\displaystyle{\Bezier{i}{n}{\alpha} (x)
=C_{n}^{i}{w}^{i}{(1-w)}^{n-i}
}$.

Since
$f_\beta \in {\cal H}{([a_1\, , \, b_1])}$ is a bijection from  $[a_1\, , \, b_1]$ into
 $[0\, , \, 1]$ then\\
 $\displaystyle{ y=f_{\beta}^{-1} (w) \in  [a_1\, , \, b_1]}$ and
$
\displaystyle{
\Bezier{i}{n}{\beta} (y)
=C_{n}^{i}{w}^{i}{(1-w)}^{n-i}
=\Bezier{i}{n}{\alpha} (x)
}$.
This implies
$$
\displaystyle{
M=B_\alpha (x) 
= \sum_{i=0}^{n} d_{i}\Bezier{i}{n}{\alpha} (x)
= \sum_{i=0}^{n} d_{i}\Bezier{i}{n}{\beta} (y)
=B_\beta (y) 
\in B_\beta
}
$$
We conclude that $\displaystyle{M \in B_\alpha \implique M \in B_\beta}$

In the same way we show that
$\displaystyle{M \in B_\beta \implique M \in B_\alpha}$

More precisely we have
$\displaystyle{
\Bezier{i}{n}{\alpha}=\Bezier{i}{n}{\beta} \circ f_{\beta}^{-1} \circ f_{\alpha}
}$
and $\forall x \in  [a\, , \, b]$
$\displaystyle{ B_\alpha (x) = B_\beta (y) }$ with
$y= f_{\beta}^{-1} \circ f_{\alpha}(x) \in  [a_1\, , \, b_1]$.

Otherwise one can observe that
$\forall f \in {\cal H}{([a\, , \, b])}$ we have $f, f^{-1}  \in {\cal C}^{\infty}{([a\, , \, b])}$.
Then we have  $f_{\beta}^{-1} \circ f_{\alpha}  \in {\cal C}^{\infty}{([a\, , \, b])}$.

For all $x \in [a\, , \, b]$ we have
$$
\begin{array}{rcl}
\displaystyle{
\frac{d B_\alpha}{dx} (x) 
}
&=&
\displaystyle{
\frac{d B_\beta}{dy} (y) 
\frac{dy}{dx}
}
\\
\displaystyle{
\frac{d^2 B_\alpha}{dx^2} (x) 
}
&=&
\displaystyle{
\frac{d^2 B_\beta}{dy^2} (y) 
{\left( \frac{d y}{dx} \right)}^2
}
+
\displaystyle{
\frac{d B_\beta}{dy} (y) 
\frac{d^2 y}{dx^2}
}\\
\end{array}
$$
with
$\displaystyle{
y=f_{\beta}^{-1} \circ f_{\alpha}(x)
}$

The curves $B_\alpha$ and  $B_\beta$ have the same tangente at the common  point 
$B_\alpha (x)$ for all  $x \in [a\, , \, b]$

Therefore
$$
\displaystyle{
\frac{d B_\alpha}{dx} (x) 
\times
\frac{d^2 B_\alpha}{dx^2} (x) 
=
\frac{d B_\beta}{dy} (y) 
\times
\frac{d^2 B_\beta}{dy^2} (y) 
{\left( \frac{d y}{dx} \right)}^3
}
$$

and
$$
\displaystyle{
\norme{
\frac{d B_\alpha}{dx} (x) 
\times
\frac{d^2 B_\alpha}{dx^2} (x) 
}=
\norme{
\frac{d B_\beta}{dy} (y) 
\times
\frac{d^2 B_\beta}{dy^2} (y) 
}
\module{
 \frac{d y}{dx}
}^3
}
$$

We deduce that $\kappa_\alpha (x)$ and  $\kappa_\beta (y)$ the respective
curvatures of $B_\alpha$ and  $B_\beta$ at the common point  
$B_\alpha (x)=B_\beta(y)$ satisty 
$$
\kappa_\alpha (x) = 
\displaystyle{
\frac{
\norme{
\frac{d B_\alpha}{dx} (x) 
\times
\frac{d^2 B_\alpha}{dx^2} (x) 
}
}{
\norme{
\frac{d B_\alpha}{dx} (x) 
}^3
}
}
=
\displaystyle{
\frac{
\norme{
\frac{d B_\beta}{dy} (y) 
\times
\frac{d^2 B_\beta}{dy^2} (y) 
}
}{
\norme{
\frac{d B_\beta}{dy} (y) 
}^3
}
}
=
\kappa_\beta (y) 
$$

This complete the proof.

\section{Conclusion}
We have an approximation basis of rational functions which unfortunately  
can not create the functions
$\displaystyle{t\mapsto \frac{t}{1+t^2}}$ 
and
$\displaystyle{t\mapsto \frac{1-t^2}{1+t^2}}$.
Thus, this new family of B\'ezier curves do not resolve the primary motivation 
of rational B\'ezier curves, which consists  of generating planar conics exactly.
This new class, however, gives another alternative construction of B\'ezier curves 
using  algorithms as soon as accurate than the standard one
with a large conservation of usual properties of Bernstein functions
 and B\'ezier curves.

The analysis of the approximation properties of  this new class of 
Bernstein functions can be of some interest.



\section*{Appendix}

\noindent
\begin{center}
Rational Bernstein basis functions of degree  2 and   derivatives :
\end{center}

$$
\Bezier{i}{2}{\alpha}(t) =
\left\lbrace 
\begin{array}{lr}
{{\left(\alpha-1\right)^2\,\left({\it b}-t\right)^
 2}\over{\left(\alpha\,{\it b}-{\it b}-\alpha\,{\it a}+t\right)
 ^2}} & \textrm{if } i=0\\
 -{{2\,\left(\alpha-1\right)\,\alpha\,\left({\it a}-t
 \right)\,\left({\it b}-t\right)}\over{\left(\alpha\,{\it b}-
 {\it b}-\alpha\,{\it a}+t\right)^2}}&\textrm{if } i=1\\
 {{\alpha^2\,\left(
 {\it a}-t\right)^2}\over{\left(\alpha\,{\it b}-{\it b}-\alpha
 \,{\it a}+t\right)^2}}&\textrm{if } i=2\\ 
 \end{array}
 \right.
 $$

$$
\frac{d}{dt} \Bezier{i}{2}{\alpha}(t) =
\left\lbrace
\begin{array}{lr}
 -{{2\,\left(\alpha-1\right)^2\,\alpha\,\left(
 {\it b}-t\right)\,\left({\it b}-{\it a}\right)}\over{\left(
 \alpha\,{\it b}-{\it b}-\alpha\,{\it a}+t\right)^3}}&\textrm{if } i=0\\
 {{2
 \,\left(\alpha-1\right)\,\alpha\,\left({\it b}-{\it a}\right)\,
 \left(\alpha\,{\it b}-{\it b}+\alpha\,{\it a}-2\,\alpha\,t+t
 \right)}\over{\left(\alpha\,{\it b}-{\it b}-\alpha\,{\it a}+t
 \right)^3}}&\textrm{if } i=1\\
 -{{2\,\left(\alpha-1\right)\,\alpha^2\,\left(
 {\it a}-t\right)\,\left({\it b}-{\it a}\right)}\over{\left(
 \alpha\,{\it b}-{\it b}-\alpha\,{\it a}+t\right)^3}}&\textrm{if } i=2\\
\end{array} 
  \right. 
$$

Rational Bernstein basis functions of degree  3 and   derivatives :
$$
\Bezier{i}{3}{\alpha}(t) =
\left\lbrace
\begin{array}{lr}
{{\left(\alpha-1\right)^3\,\left({\it b}-t
 \right)^3}\over{\left(\alpha\,{\it b}-{\it b}-\alpha\,{\it a}+
 t\right)^3}}
&\textrm{if } i=0\\ 
 -{{3\,\left(\alpha-1\right)^2\,\alpha\,\left(
 {\it a}-t\right)\,\left({\it b}-t\right)^2}\over{\left(\alpha\,
 {\it b}-{\it b}-\alpha\,{\it a}+t\right)^3}}
&\textrm{if } i=1\\
 {{3\,
 \left(\alpha-1\right)\,\alpha^2\,\left({\it a}-t\right)^2\,\left(
 {\it b}-t\right)}\over{\left(\alpha\,{\it b}-{\it b}-\alpha\,
 {\it a}+t\right)^3}}
&\textrm{if } i=2\\
 -{{\alpha^3\,\left({\it a}-t
 \right)^3}\over{\left(\alpha\,{\it b}-{\it b}-\alpha\,{\it a}+
 t\right)^3}} 
&\textrm{if } i=3\\
\end{array} 
 \right.
 $$

 $$
\frac{d}{dt} \Bezier{i}{3}{\alpha}(t) =
\left\lbrace
\begin{array}{lr}
 -{{3\,\left(\alpha-1\right)^3\,\alpha\,\left(
 {\it b}-t\right)^2\,\left({\it b}-{\it a}\right)}\over{\left(
 \alpha\,{\it b}-{\it b}-\alpha\,{\it a}+t\right)^4}}
&\textrm{if } i=0\\ 
 {{3\,\left(\alpha-1\right)^2\,\alpha\,\left({\it b}-t\right)\,
 \left({\it b}-{\it a}\right)\,\left(\alpha\,{\it b}-{\it b}+
 2\,\alpha\,{\it a}-3\,\alpha\,t+t\right)}\over{\left(\alpha\,
 {\it b}-{\it b}-\alpha\,{\it a}+t\right)^4}}
&\textrm{if } i=1\\
 -{{3\,
 \left(\alpha-1\right)\,\alpha^2\,\left({\it a}-t\right)\,\left(
 {\it b}-{\it a}\right)\,\left(2\,\alpha\,{\it b}-2\,{\it b}+
 \alpha\,{\it a}-3\,\alpha\,t+2\,t\right)}\over{\left(\alpha\,
 {\it b}-{\it b}-\alpha\,{\it a}+t\right)^4}}
&\textrm{if } i=2\\
 {{3\,
 \left(\alpha-1\right)\,\alpha^3\,\left({\it a}-t\right)^2\,\left(
 {\it b}-{\it a}\right)}\over{\left(\alpha\,{\it b}-{\it b}-
 \alpha\,{\it a}+t\right)^4}}
&\textrm{if } i=3\\
\end{array} 
 \right.
 $$

\end{document}